\documentclass[a4paper,11pt]{article}
\pdfoutput=1 

\usepackage{jheppub} 
\usepackage{physics}
\usepackage{subfigure}
\usepackage{cancel}

\usepackage[T1]{fontenc}

\title{\boldmath Gravitational collapse of quantum fields and Choptuik scaling}

\author[a]{Benjamin Berczi,}
\author[a]{Paul M. Saffin,}
\author[b,c]{and Shuang-Yong Zhou}

\affiliation[a]{School of Physics and Astronomy, University of Nottingham, University Park, Nottingham NG7 2RD, United Kingdom}
\affiliation[b]{Interdisciplinary Center for Theoretical Study, University of Science and Technology of China, Hefei, Anhui 230026, China}
\affiliation[c]{Peng Huanwu Center for Fundamental Theory, Hefei, Anhui 230026, China}

\emailAdd{benjamin.berczi@nottingham.ac.uk}
\emailAdd{paul.saffin@nottingham.ac.uk}
\emailAdd{zhoushy@ustc.edu.cn}

\abstract{Gravitational collapse into a black hole has been extensively studied with classical sources. We develop a new formalism to simulate quantum fields forming a black hole. By choosing a convenient coherent state, this formalism taps into well-established techniques used for classical collapse and adds on the evolution of the mode functions of the quantum field operator. Divergences are regularized with the cosmological constant and Pauli-Villars fields. Using a massless spherically symmetric scalar field as an example, we demonstrate the effectiveness of the formalism by reproducing some classical results in gravitational collapse, and identifying the difference due to the quantum effects. We also find that Choptuik scaling in critical collapse survives in the semiclassical simulation, and furthermore the quantum deviation from the classical Choptuik scaling decreases when the system approaches the critical point.}

\preprint{\footnotesize USTC-ICTS/PCFT-21-44}

\begin{document} 
\maketitle
\flushbottom

\section{Introduction}\label{sec:intro}

Black holes are fascinating objects in the universe that form as the end states of gravitational collapse, a unique manifestation of the strength and special nature of gravity. Since Hawking's famous calculation \cite{hawkingradiation}, we know that black holes have a temperature, as they can evaporate via  Hawking quanta. This is an astonishing result, as it connects general relativity to quantum field theory, which is another pillar of modern physics that underlies all known matter sources in the universe. While stationary astronomical black holes are expected to exhibit little in the way of quantum effects outside the event horizons, the dynamical formation of black holes or other compact stars may involve extreme matter sources that are quantum in essence. Moreover, primordial black holes may form in the early universe \cite{Carr:2020gox} and can have a high temperature and large Hawking radiation. Purely theoretically, black holes along with their quantum properties have caused great puzzles but also great advances in our understanding of gravity and quantum field theory.

The dynamics of classical gravitational collapse with various matter sources have been studied over the years. The simple pressure-less homogenous dust model was studied as early as the late 1930s \cite{Oppenheimer:1939ue,Datt}. This has been generalized to spherically symmetric inhomogeneous dust models \cite{Eardley:1978tr, Christodoulou:1984mz,Newman:1985gt, Waugh:1988ud, 
Joshi:1993zg, Singh:1994tb} and more general perfect fluid models \cite{Ori:1987hg, Ori:1989ps, Joshi:1992vr, Dwivedi:1994qs, Harada:1998wb,
Lake:1992zz}.
While black holes are the natural end states of gravitational collapse, it is singularities that must form as the end states of gravitational collapse from reasonable matter sources, as required by the singularity theorems of Hawking and Penrose \cite{Hawking:1970zqf,Penrose:1964wq}. The cosmic censorship conjecture \cite{Penrose:1969pc} suggests that naked singularities should be cloaked by horizons under appropriate assumptions so as to avoid loss of predictability at the singularities. However, naked singularities do appear in the above models of gravitational collapse. A major purpose of the gravitational collapse studies is to clarify the scopes and the assumptions of the cosmic censorship conjecture. See \cite{Joshi:2011rlc} for a review on gravitational collapse and spacetime singularities.

 At a more fundamental level, one may model the matter source as a scalar field, which provides arguably the simplest arena to probe the nonlinearities of general relativity and its interaction with matter. Classical scalar collapse in spherical symmetry has been  analytically investigated by Christodoulou \cite{christodoulou1, christodoulou2, christodoulou3, christodoulou4, Christodoulou:1994hg, Christodoulou:1999}, which also provides counter-examples to the naive cosmic censorship \cite{Christodoulou:1994hg}. Indeed, as numerically uncovered by Choptuik \cite{choptuik}, interesting critical phenomena exist in spherical scalar collapse that are characterized by scale invariance and universal scaling of the black hole mass with initial data. The critical collapse happens at the boundary between the parameter region of regular initial data that can form black holes and the region that can not, and the singularity at critical collapse is exactly a naked singularity, as explicitly shown in an an analytical example \cite{Brady:1994aq}. \cite{Gundlach:1995kd, Gundlach:1996eg} provided some further insights into the critical phenomena. Critical collapse has also been observed with axisymmetric gravitational waves \cite{Abrahams:1993wa}, spherical collapse of radiation fluid \cite{Evans:1994pj} and a non-Abelian field \cite{Choptuik:1996yg}. A real analytical solution of the Choptuik spacetime has recently been constructed \cite{Reiterer:2012hnr}.
See \cite{Gundlach:2007gc} for a review of critical collapse.

In this paper, we study the semiclassical gravitational collapse of a massless scalar in spherical symmetry. That is, we shall keep the metric field at the classical level but fully quantise the scalar field. Such semiclassical scalar collapse is a demanding task, early attempts along this line of enquiry include \cite{tomimatsu, Bak:1999wb, Bak:2000kg}. The quantum properties of spherical shell collapse have also been investigated by \cite{Vachaspati:2007hr, Greenwood:2008ht, Vachaspati:2018pps}, and \cite{Russo:1992ax, Strominger:1993tt, Piran:1993tq, Lowe:1992ed} studied dilaton collapse in 2D gravity, which is drastically different from a 4D model for a number of reasons. 
 
We will present a formalism that allows us to simulate the gravitational collapse of a quantum scalar field into a black hole. This is the companion paper of \cite{Berczi:2020nqy} in which we reported the use of this formalism to obtain some initial results. (Ref \cite{Guenther:2020kro}, which appeared after \cite{Berczi:2020nqy}, studied a similar system with a different method that only included the spherical mode for the quantum field.) In the present paper we provide a more thorough overview of the formalism with all essential technical details. Specifically, we couple a spherically symmetric massless quantum scalar field to classical gravity. Importantly, the notion of coherent states is utilised so that we can build the semiclassical system upon the previously well-studied classical collapse. Namely, the expectation value of the quantum scalar field operator in the coherent state is identified with the classical scalar field of classical collapse. This is a crucial point in the formalism because it allows us to split the stress-energy tensor components into a sum of contributions from the "classical field" and contributions from quantum modes. 

It is also essential in our formalism to make use of the latest stable numerical relativity formulations of the Einstein equations, as numerically the quantum field simulation corresponds to a simulation of thousands of mode functions, which is computationally much more demanding in numerical accuracy and stability than the classical collapse. Also, to further sustain a sufficiently long period of simulation, we use as high as a tenth order finite difference method and introduce artificial dissipation terms in our equations of motion. Another major difference in simulations with products of quantum fields evaluated at the same spacetime point is that there will be UV divergences that needs cancelling, similar to loop calculations in quantum field theory. We find that the Pauli-Villars regularisation with five fields plus a cosmological constant is sufficient for our purposes.

We reproduce the results of classical scalar collapse to a black hole with our semi-classical simulations and extract the quantum deviations for the stress-energy tensor. We also simulate the Choptuik scaling for the semiclassical critical collapse and find that when approaching the critical point the quantum deviations from the classical Choptuik scaling actually decrease. We validate our results with various convergence studies.

The paper is organised as follows. In Section \ref{section2} the classical collapse of a spherically symmetric scalar field is described, which will provide a basis for the quantum system. The evolution of both matter fields and metric fields are described in detail. In Section \ref{section3} the aforementioned classical scalar field is quantised and expanded in terms of spherically symmetric mode functions. The chosen quantum state for the system is described, along with the expectation values of quantities relevant for calculation of the evolution equations of the geometry. We also describe how to regularise the system, since it has divergences that must be dealt with. In Section \ref{section5} the numerical methods used to run the simulation and how the quantum vacuum is built up are discussed. Then, in Section \ref{section6}, we present the results about the evolution of the quantum collapse into a black hole, highlighting the quantum corrections to the classical collapse, and also the results on the quantum Choptuik scaling near critical collapse. Lastly, we provide the convergence studies of our simulations in Section \ref{sectionconv}, and conclude in Section \ref{sectionsum}.

\section{Classical collapse} \label{section2}
In this section the theory for the simulation of the collapse of a spherically symmetric classical massless scalar field is discussed using the methods of \cite{alcubierre} but with a different gauge choice. First, the geometrical side of the Einstein equation will be presented using variables that make the evolution equations better suited for lattice simulations, and then we show how to include the matter fields. Before connecting the geometry to the matter equations, the gauge choices are reviewed briefly. Lastly, the choices for the initial conditions for the simulation are shown. Note that in the simulation all the fundamental constants are set to one, $\hbar=c=M_P=1$. If displayed, their purpose is simply illustration.

\subsection{Geometry}

We will restrict our simulations to have spherical symmetry in $3+1$D in this paper. For classical simulations, it is certainly viable to do full $3+1$D simulations with modern hardware, but real time simulations with quantum fields involve solving partial differential equations for many quantum modes simultaneously, which is computationally very expensive. For our purposes, the following spherically symmetric line element is chosen:
\begin{equation}
\label{metric0}
    ds^2=-\alpha^2(t,r)dt^2+A(t,r)dr^2+r^2B(t,r)(d\theta^2+\sin^2\theta d\varphi^2), 
\end{equation}
where $\alpha(t,r)$ is the lapse function and $A(t,r)$ and $B(t,r)$ are functions governing the spatial part of the metric. The lapse function $\alpha(t,r)$ has a crucial role to play in the collapse --- not only will it be used to choose a Bona-Masso type gauge condition, but also one can use it to identify the formation of a black hole, due to the ``collapse of the lapse'' phenomenon \cite{alcubierre}.

However, we shall not directly use $\alpha(t,r)$, $A(t,r)$ and $B(t,r)$ as our dynamical variables, with which the Einstein equations are second order differential equations. To cast the Einstein equations as first order equations, we can define the following variables
\begin{equation}
\begin{split}
    K_A&=-\frac{1}{2\alpha}\frac{\dot{A}}{A}, \: \quad \quad D_A=\frac{A'}{A}, \\
    K_B&=-\frac{1}{2\alpha}\frac{\dot{B}}{B}, \: \quad \quad D_B=\frac{B'}{B}, \\
    \lambda&=\frac{1}{r}\Big(1-\frac{A}{B} \Big), \quad D_{\alpha}=\frac{\alpha'}{\alpha} ,
\end{split}
\end{equation}
where the dot represents differentiation with respect to $t$, and the prime represents differentiation with respect to $r$.

Note that $\lambda$ is defined in order to guarantee the local flatness of the metric (at $r=0$). Local flatness in these coordinates manifests itself as $A(t, r=0)=B(t,r=0)$ for all $t$. Many $1/r$ terms in the evolution equations discussed below are only regular (not divergent) numerically, if this local flatness condition is exactly met. Small numerical errors in $A(t, r=0)$ and $B(t,r=0)$ could mean exponential divergences in the numerical system. Hence, using the definition of $\lambda(t,r)$, and further requiring it to be antisymmetric, the local flatness condition is exactly met at each time step. 

In addition, to ensure that the system of equations is strongly hyperbolic for any gauge choice of $\alpha(t,r)$, we must perform some further changes of variables. Namely, instead of $(K_A, D_A)$, we will use $(K, \Tilde{U})$ respectively,
\begin{equation}
\begin{split}
    K&=K_A+2K_B, \\
    \Tilde{U}&=D_A-2D_B-\frac{4B\lambda}{A}.
\end{split}
\end{equation}
The detailed motivation for these choices are explained more in depth in \cite{alcubierre}, but essentially for some gauge choices for the lapse function $\alpha(t,r)$ the original system is not strongly hyperbolic. Hence, using the above change of variables, our system is strongly hyperbolic for any gauge choice of $\alpha(t,r)$, which are discussed at the end of this subsection.

Then the non-zero components of the Einstein tensor can be written as
\begin{align}
    G^t_{\:\:\: r}=&\frac{2}{\alpha}\Bigg[ -K_B'+\Big( \frac{1}{r} +\frac{D_B}{2} \Big)(K-3K_B) \Bigg], \label{Grt} \\
    G^t_{\:\:\: t}=&\frac{1}{A}\Bigg[ D_B'+\frac{1}{r}\Big( \lambda +D_B-\Tilde{U}-\frac{4\lambda B}{A} \Big)
    -D_B\Big( \frac{1}{4}D_B+\frac{1}{2}\Tilde{U}+\frac{2\lambda B}{A}\Big) \Bigg] \notag \\
    &-K_B(2K-3K_B), \label{Gtt}\\
    G^r_{\:\:\: r}=&\frac{2}{\alpha}\Bigg[\dot{K}_B-\frac{3}{2}\alpha K_B^2 +\frac{\alpha}{2r^2A}\Big( 1-\frac{A}{B} \Big) +\frac{\alpha}{2rA}D_B +\frac{\alpha}{8A}D_B^2 \notag \\
    &+\frac{\alpha}{rA}D_{\alpha}+\frac{\alpha}{2A}D_B D_{\alpha} \Bigg],\\
    G^{\theta}_{\:\:\: \theta}=&\frac{1}{\alpha}(\dot{K}-\dot{K}_B) 
    -K^2+3K_B(K-K_B) +\frac{1}{2A}(D_B'+2D_{\alpha}') \notag \\
    &-\frac{1}{4A}\Bigg[(D_B+2D_{\alpha})\Big( 
    \Tilde{U}+D_B+\frac{4B\lambda}{A}\Big)-4D_{\alpha}^2 \Bigg] \notag \\
    &+ \frac{1}{2rA}\Big( 2D_{\alpha}-\Tilde{U}-\frac{4B\lambda}{A} \Big).  
\end{align}
The $G^r_{\:\:\: t}$ and $G^t_{\:\:\: t}$ components of the Einstein equations are actually constraint equations, which do not contain time derivatives with the new dynamical variables above. They are called the momentum and Hamiltonian constraint respectively
\begin{align} 
\label{momentum}
    P= &K_B'-\Big( \frac{1}{r} +\frac{D_B}{2} \Big)(K-3K_B) + \frac{\alpha}{2}G^t_{\:\:\: r}=0, \\
\label{Hcondef}
    H= &D_B'+\frac{1}{r}\Big( \lambda +D_B-\Tilde{U}-\frac{4\lambda B}{A} \Big)
    -D_B\Big( \frac{1}{4}D_B+\frac{1}{2}\Tilde{U}+\frac{2\lambda B}{A}\Big) \notag \\ 
    &- AK_B(2K-3K_B)-AG^t_{\:\:\: t}=0.  
\end{align}
where $G^t_{\:\:\: r}$ and $G^t_{\:\:\: t}$ are understood to be replaced with the corresponding components of the stress-energy tensor via the Einstein equations. These constraints provide a sanity check on whether numerical errors are under control in the simulation.

Our metric ansatz, Eq.~\eqref{metric0}, contains gauge/coordinate degrees of freedom. Particularly, we may freely choose $\alpha$ as a gauge choice, which slices the spacetime into hypersurfaces with the same times. Since we are dealing with gravitational collapse, we need to choose a singularity avoiding gauge choice. This ensures that even when a singularity is present inside an apparent horizon, the spatial grid outside of the horizon may still be evolved. The two established families of singularity avoiding gauge choices are maximal slicing conditions and Bona-Masso type slicing conditions \cite{BonaMasso}.

The maximal slicing conditions are used in \cite{alcubierre}, which is equivalent to demanding that the extrinsic curvature of the equal time hypersurface is constant in time. This leads to a constraint equation involving $\alpha$ that needs to be integrated with respect to $r$ in spherical symmetry. 

The Bona-Masso type slicing conditions \cite{BonaMasso}, on the other hand, treat $\alpha$ as a dynamical variable, which is artificially specified by an evolution equation along with the evolution of other physical variables. These are also called hyperbolic slicing conditions, since $\alpha(t,r)$ obeys a wave equation with a source containing $K$.

Although maximal slicing is a robust gauge choice, it is computationally expensive and slow, since one needs to integrate spatially at every time step. The Bona-Masso type gauges possess similar robustness, but requires less computational power, since one just has an additional dynamical field. Therefore, in our simulations, we use the Bona-Masso slicing conditions. Specifically, the lapse function $\alpha(t,r)$ is chosen to satisfy the following evolution equation
\begin{equation}
    \dot{\alpha}=-\alpha^2 f(\alpha)K,
\end{equation}
where $f(\alpha)$ is a suitable function of $\alpha$, to be tuned in a particular problem. The $\alpha$ evolution also leads to the evolution of $D_{\alpha}$, which is given by
\begin{equation}
    \dot{D}_{\alpha}=-\partial_r(\alpha f(\alpha)K).
\end{equation}
The most used gauge choice within the Bona-Masso family is the 1+log gauge, which is given by
\begin{equation}
    f(\alpha)=\frac{2}{\alpha}.
\end{equation}
This choice ensures strong singularity avoidance and is the one we use.

\subsection{Matter}
Let us now introduce the scalar field involved in the simulation. We shall for simplicity work with a free massless scalar $\Phi$, which couples minimally to gravity. Classically, the scalar field acts as a stress-energy source to the Einstein equation, and gravity also influences the matter via the Klein-Gordon equation. We have the following stress-energy tensor for the scalar
\begin{equation}
    T_{\mu \nu} = \partial_{\mu}\Phi \partial_{\nu}\Phi 
                    -\frac{1}{2} g_{\mu \nu}\Big[g^{\rho\sigma} \partial_{\rho}\Phi \partial_{\sigma}\Phi \Big] .
\end{equation}
We shall introduce the following new field variables for the scalar field
\begin{equation}
    \Pi=\frac{A^{\frac{1}{2}}B}{\alpha}\dot{\Phi}, ~~~~~~~~
    \Psi=\Phi' ,
\end{equation}
which are needed to formulate the scalar evolution equations with only first order derivatives. With these variables, the stress-energy tensor can be decomposed into
\begin{align}
\label{rhodef}
    \rho&=n^{\mu}n^{\nu}T_{\mu\nu}=\frac{1}{2A}\Bigg( \frac{\Pi^2}{B^2}+\Psi^2 \Bigg) + \frac{1}{2Br^2}(\partial_{\theta}\Phi)^2 + \frac{1}{2Br^2 \sin^2 \theta}(\partial_{\varphi}\Phi)^2, \\
    j_A&=-n^{\mu}T_{\mu r}=-\frac{\Pi \Psi}{A^{\frac{1}{2}}B}, \\
    S_A&=\gamma^{rr}T_{rr}=\frac{1}{2A}\Bigg( \frac{\Pi^2}{B^2} +\Psi^2 \Bigg) - \frac{1}{2Br^2}(\partial_{\theta}\Phi)^2 - \frac{1}{2Br^2\sin^2 \theta}(\partial_{\varphi}\Phi)^2, \\
    S_B&=\gamma^{\theta\theta}T_{\theta\theta} = \frac{1}{2A}\Bigg( \frac{\Pi^2}{B^2} - \Psi^2 \Bigg) + \frac{1}{2Br^2}(\partial_{\theta}\Phi)^2 - \frac{1}{2Br^2\sin^2 \theta}(\partial_{\varphi}\Phi)^2.
    \label{SBdef}
\end{align}
Here $n^{\mu}$ is the unit timelike vector orthogonal to the equal time hypersurface, and $\gamma^{ij}$ is the inverse of the spatial part of the metric. Note that in our coordinates these quantities are simply $\rho=-T^t{}_t$, $j_A=T^t{}_r$, $S_A=T^r{}_r$ and $S_B=T^{\theta}{}_{\theta}$. 
\subsection{Evolution equations}
The Einstein equations and the scalar equation of motion are respectively given by
\begin{align}
    G_{\mu \nu} &= \frac{1}{M_P^2} \big(T_{\mu \nu}-\Lambda g_{\mu \nu}\big), \\
    \Box \Phi &= 0,
    \label{kgequ}
\end{align}
where $M_P=1/\sqrt{8\pi G}$ is the reduced Planck mass and we have also added the cosmological constant $\Lambda$ as a stress-energy source. For classical simulations, we can set the  cosmological constant to zero,  but the presence of a cosmological constant is required to regularise the quantum field in the semiclassical system, as will be explained in Section \ref{regularisation} in detail. 

As mentioned above, we want to cast these field equations in a first order form, which can be achieved by evolving the following variables introduced previously
\begin{equation}
    (\Phi^{(s)}, \: \Psi^{(a)},\: \Pi^{(s)},\: A^{(s)},\: B^{(s)},\: D_B^{(a)},\: \Tilde{U}^{(a)},\: K^{(s)},\: K_B^{(s)},\: \lambda^{(a)},\: \alpha^{(s)},\: D_{\alpha}^{(a)}),
\end{equation}
where the first three fields are matter fields, the others are metric fields, and the (temporarily added) superscript indicates whether the field is symmetric or antisymmetric around $r=0$, as we shall be using initial data with definite parity to help with numerical stability. The Klein-Gordon equation (\ref{kgequ}) can be cast in the following form
\begin{align} 
    \dot{\Phi}&=\frac{\alpha}{A^{\frac{1}{2}}B}\Pi, \\
    \dot{\Psi}&=\partial_r\Bigg(\frac{\alpha}{A^{\frac{1}{2}}B}\Pi \Bigg), \\
    \dot{\Pi}&=\frac{1}{r^2}\partial_r\Bigg(\frac{\alpha B r^2}{A^{\frac{1}{2}}}\Psi \Bigg)
    +\frac{\alpha A^{\frac{1}{2}}}{r^2}\Big[ \frac{1}{\sin{\theta}}\partial_{\theta}(\sin{\theta}\partial_{\theta}\Phi)+\frac{1}{\sin{\theta}^2}\partial_{\varphi}\partial_{\varphi}\Phi \Big]. \label{pi_evolution}
\end{align}
As for the metric fields, the relevant components of the Einstein equations are the following:
\begin{align} 
    G^r_{\:\:\: t}&=-\frac{\alpha j_A}{AM_P^2}, \label{einsteincomponent1}\\
    G^t_{\:\:\: t}&=-\frac{1}{M_P^2}\big( \rho + \Lambda \big), \label{einsteincomponent2}\\
    G^r_{\:\:\: r}&=\frac{1}{M_P^2}\big( S_A - \Lambda \big), \label{einsteincomponent3}\\
    G^{\theta}_{\:\:\: \theta}&=\frac{1}{M_P^2} \big( S_B- \Lambda \big). \label{einsteincomponent4} 
\end{align}
where the first two equations are constraint equations, {\it i.e.,} Eq.~\eqref{momentum} and Eq.~\eqref{Hcondef}, and the latter two are evolution equations of our chosen dynamical fields. For $A(t,r),\;B(t,r)$ and $D_B(t,r)$, their evolution equations can be found straightforwardly by simply using their definitions:
\begin{align}
    \dot{A}=&-2\alpha A (K-2K_B), \\
    \dot{B}=&-2\alpha B K_B, \\
    \dot{D}_B=&-2\partial_r(\alpha K_B).
\end{align}
The evolution of the gauge degrees of freedom $\alpha(t,r)$ and $D_{\alpha}(t,r)$ are to be chosen by hand, that is, by imposing the following slicing conditions
\begin{align}
    \dot{\alpha}=&-\alpha^2 f(\alpha) K, \\
    \dot{D}_{\alpha} =& -\partial_r(\alpha f(\alpha)K), 
\end{align}
We will work in 1+log gauge throughout, for which we choose $f(\alpha)={2}/{\alpha}$. To find the evolution equations of $K(t,r)$ and $K_B(t,r)$, the Einstein equations have to be used. The equation of motion for $K_B(t,r)$ is derived by combining the $G_{\:\:\: r}^r$ and $G_{\:\:\: t}^t$ components, which gives
\begin{equation}
\begin{split}
    \dot{K}_B&=\frac{\alpha}{Ar}\Big[ \frac{1}{2}\Tilde{U}+\frac{2\lambda B}{A} -D_B-\lambda-D_{\alpha} \Big]  \\ 
    &~~~ + \frac{\alpha}{A} \Big[
    -\frac{1}{2}D_{\alpha}D_B-\frac{1}{2}D_B'+\frac{1}{4}D_B\Big( \Tilde{U}+\frac{4\lambda B}{A} \Big)+AKK_B \Big]  \\
    &~~~ + \frac{\alpha}{2M_P^2}(S_A-\rho-\Lambda).
\end{split}
\end{equation}
For the evolution of $K$, the equations for $G_{\:\:\: r}^r$, $G_{\:\:\: t}^t$ and $G_{\:\:\: \theta}^{\theta}$ can be used to find
\begin{equation}
\begin{split}
    \dot{K}&=\alpha(K^2-4KK_B+6K_B^2)  \\
    &~~~ -\frac{\alpha}{A}\Big[
    D_{\alpha}'+D_{\alpha}^2+\frac{2D_{\alpha}}{r}-\frac{1}{2}D_{\alpha}
    \Big( \Tilde{U} + \frac{4\lambda B}{A} \Big) \Big]   \\
    &~~~ +\frac{\alpha}{2M_P^2}(\rho + S_A + 2S_B-2\Lambda). \\ 
\end{split}
\end{equation}
For $\lambda$, using its definition, we can get
\begin{equation} \label{lambdadot}
    \dot{\lambda}= \frac{2\alpha A}{B}\frac{(K-3K_B)}{r},
\end{equation}
which is well defined in the continuum limit. However, the ${1}/{r}$ factor on the right hand side makes this evolution equation numerically highly unstable when discretising the equation and putting it on a lattice. To overcome this problem, the momentum constraint (\ref{momentum}) can be used to remove this factor and we arrive at
\begin{equation}
    \dot{\lambda}=\frac{2\alpha A}{B}\Big[ K_B'-\frac{1}{2}D_B(K-3K_B)+
    \frac{j_A}{2M_P^2} \Big]. 
\end{equation}
Note that other $1/r$ terms in the evolution equations (such as in Eq.~\eqref{pi_evolution}) are automatically regular by demanding the dynamical variables to be symmetric or antisymmetric. However here, in Eq.~\eqref{lambdadot}, the combination of $(K-3K_B)/r$ must be regular. Analytically, this is true since the local flatness condition $A(t,r=0)=B(t,r=0)$ means also that $K_A(t,r=0)=K_B(t,r=0)$, hence $K(t,r=0)=3K_B(t,r=0)$. Numerically, on the other hand, small errors can spoil this cancellation hence making the term unstable. Thus the need to make it regular by using the momentum constraint. 

Similarly, using its definition and the Hamiltonian constraint Eq.~\eqref{Hcondef}, the evolution for $\Tilde{U}$ can be found to be
\begin{equation}
\begin{split}
    \dot{\Tilde{U}}=&-2\alpha \Big[K'+D_{\alpha}(K-4K_B)-2(K-3K_B)\Big( D_B-\frac{2\lambda B}{A} \Big)\Big]  \\
    &-4\alpha \frac{j_A}{M_P^2}. \\ 
\end{split}
\end{equation}
This completes the derivation of the evolution equations for our dynamical variables, which, unlike the original Einstein equations or even the ADM decomposed version, are generally well behaved numerically after discretisation.

\subsection{Initial conditions} \label{initialcondsclass}
To evolve the system, some initial conditions must be chosen. Some care must be taken for a system with gauge degrees of freedom, in which case some of the equations of motion are actually constraint equations. We follow a free evolution scheme in which the constraints are solved once initially when preparing the initial data. Then the keeping of the constraint equations can be used to monitor how well the simulation works.

 For the matter fields, we choose the initial conditions to be
\begin{equation}
\label{PhiPsi0}
    \Phi^0=f(r), ~~~
    \Psi^{0}=f'(r), ~~~
    \Pi^0=0. 
\end{equation}
where here the superscript $0$ denotes the initial condition of the corresponding quantity at $t=0$. A natural choice for our purposes is to take the function $f(r)$ to be a Gaussian wave packet. In our simulations, we use the following family of functions
\begin{equation} \label{initialfield}
    f(r) = a\exp\left[-(r/D)^2\right],
\end{equation}
where $a$ is the amplitude and $D$ the width of the wave packet. Note that the above definition of $f(r)$ ensures that the initial scalar field is an even function, and thus all initial $1/r$ terms in the equations of motion are regular.
The initial conditions for the gravity fields are chosen to be 
\begin{equation}
\begin{split}
    K^0 &=K_B^0=0, \\
    \alpha^0 &=B^0=1, \\
    D_{\alpha}^0 &=D_B^0=0 
\end{split}
\end{equation}
such that they solve the constraint equations.  Note that the choice $\Pi^0=0$ is convenient, since then we have $j_A^0=0$, which means that  the momentum constraint in Eq.~\eqref{momentum} is automatically solved by the choices $K^0=K_B^0=0$. The initial conditions for the other gravity fields require some calculations. To do this, we first take $A^0(r=0)=1$, and integrate out the following equation to find the $A^0(r)$ function,
\begin{equation}
    \partial_rA^0=A^0 \Big( \frac{1}{r}(1-A^0) + \frac{r}{2M_P^2}(\Psi^0)^2 + \frac{rA^0}{M_P^2}\Lambda \Big),
\end{equation}
which comes from the Hamiltonian constraint along with the initial condition $\rho^0 =  (\Psi^0)^2/{(2A^0)}$. Then to find $\lambda^0$ and $\Tilde{U}^0$, we use their definitions along with $A^0$:
\begin{equation}
\begin{split}
    \lambda^0&=\frac{1}{r}\Big( 1-\frac{A^0}{B^0} \Big), \\
    \Tilde{U}^0&=\frac{A^{0}{}'-4\lambda^0}{A^0}. \\
\end{split}
\end{equation}
Thus we have initial conditions for all classical dynamical variable fields. This completes the setup for the classical gravitational collapse in spherical symmetry, which provides the base framework for the quantum/semi-classical simulations. We will run the classical simulations along with the semi-classical ones for a comparison later. Before that, in the following, we shall first set up our formalism to simulate quantum fields on curved space.

\section{Quantisation} \label{section3}
We want to develop a semiclassical simulation scheme where the metric fields are still treated classically but we can consistently treat the matter fields as quantum fields. This is a valid approach in situations where the quantum effects of the matter fields have become important but the quantum gravity effects have yet to kick in. For our case, it is only the classical scalar field that needs to be promoted to be a quantum scalar field. To this end, the c-number $\Phi$ needs to be upgraded to an operator $\hat\Phi$, and its equation of motion is treated as an operator equation. We can not directly simulate the operator equation, but once we expand $\hat\Phi$ in terms of creation and annihilation operators, the $\hat\Phi$ operator equation can be decomposed into the evolution equations for  the functions in front of the creation and annihilation operators, the mode functions, which are c-numbers and can be put in a lattice. Of course, the difficulty is that there are infinitely many of these mode functions. Since the classical simulation is spherically symmetric, it is also convenient to further expand the mode functions in terms of the spherical harmonics $Y(\theta, \varphi)$. The feedback of the quantum scalar field to the classical geometry is obtained via the semiclassical Einstein equations, by taking the quantum expectation value of the stress-energy tensor of the scalar field. Explicitly, we will work with the following semiclassical system of equations of motion
\begin{align}
\label{Gqeq}
G_{\mu\nu} &= \frac{1}{M_P^2} \Big[ \bra{\chi} \hat{T}_{\mu\nu} \ket{\chi} - \Lambda g_{\mu\nu}  \Big],
\\
\label{Phiqeq}
\Box \hat \Phi & =0,
\end{align}
where $\hat T_{\mu \nu} = \partial_{\mu}\hat\Phi \partial_{\nu}\hat\Phi -\frac{1}{2} g_{\mu \nu}\big[g^{\rho\sigma} \partial_{\rho}\hat\Phi \partial_{\sigma}\hat\Phi \big]$ and a coherent state will be chosen for the quantum state $\ket{\chi}$.

The key element of our formalism is that, due to a convenient choice of the quantum coherent state, the sources for the Einstein equation can be nicely separated into a classical background plus some quantum fluctuations. This fact allows us to utilise all the existing classical numerical relativity setup, reviewed in the previous section, and simply add the quantum mode functions of the matter fields on top of the simulation of the classical fields. Crucially, though, the scalar field remains fully quantum mechanical.

\subsection{Quantum scalar field}
As mentioned, the real scalar field is now promoted to a Hermitian quantum field operator $\hat\Phi$, which can be expanded in spherical mode functions as
\begin{equation}
\label{mode_expansion}
    \hat{\Phi}=\sum_{l,m} \int dk \Big[\hat{a}_{k,l,m}\Tilde{u}_{k,l}(t,r)Y^m_l(\theta, \varphi)+ \hat{a}^{\dag}_{k,l,m}\Tilde{u}^{*}_{k,l}(t,r)Y^{m*}_l(\theta, \varphi)\Big],
\end{equation}
where $Y^m_l(\theta, \varphi)$ are the spherical harmonics and in the following we shall mostly use the mode functions with a factor of $r^l$ stripped off
\begin{equation}
     u(t,r) \equiv \frac{\Tilde{u}(t,r)}{r^l} .
\end{equation}
For notational simplicity, we have suppressed the $k,l$ indices for $u$ and $\Tilde u$, and we will also suppress these indices for the $\pi$ and $\psi$ quantities defined shortly, when not confusing. The $u(t,r)=u_{k,l}(t,r)$'s are the unknown quantities to be solved in our simulations.  The operators $\hat{a}_{k,l,m}$ and $\hat{a}^{\dag}_{k,l,m}$ are the corresponding ladder operators for each quantum mode denoted by the indices $k,l,m$. The standard route to quantisation is then to impose the canonical commutation relation between the scalar field and its conjugate momentum
\begin{equation}
\begin{split}
    \Big[ \hat{\Phi}(t,\Vec{x}), \hat{\Pi}(t,\Vec{x}') \Big] = i\hbar\delta(\Vec{x}-\Vec{x}').
\end{split}
\end{equation}
If we require the mode functions satisfy
\begin{equation}
i\int dr r^2\big[ \Tilde u_{k,l}^* \partial_t \Tilde u_{k',l'} -\partial_t \Tilde u_{k,l}^* \Tilde u_{k',l'} \big] = \delta(k-k')\delta_{ll'}
\end{equation}
this is equivalent to requiring the ladder operators to obey
\begin{equation}
    \Big[\hat{a}_{k,l,m}, \hat{a}^{\dag}_{k',l',m'}\Big] = \hbar c^2 \delta_{ll'}\delta_{mm'} \delta(k-k').
\end{equation}
We want to emphasize that although the classical geometry is restricted to spherical symmetry, the quantum operators are allowed to fluctuate non-spherical symmetrically. The spherical symmetry is reflected in the quantum state we will choose below.

Similar to the classical scalar field, we also define the following variables for the quantum mode functions in order to formulate the evolution equations in a first order form
\begin{equation}
    \pi=\frac{A^{\frac{1}{2}}B}{\alpha}\dot{u}, ~~~~
    \psi=u'.
\end{equation}
We then note that the evolution equations for the mode functions $u$ and $\psi$ are just
\begin{align}
    \dot{{u}}&=\frac{\alpha}{A^{\frac{1}{2}}B}\pi, \\
    \dot{\psi}&=\partial_r\Bigg(\frac{\alpha}{A^{\frac{1}{2}}B}\pi \Bigg). 
\end{align}
For the $\pi$ evolution equation, one can use Eq.~\eqref{mode_expansion} along with Eq.~\eqref{pi_evolution} to find that
\begin{equation} \label{pi mode evolution}
    \dot{\pi}=\partial_r \Bigg( \frac{\alpha B}{A^{\frac{1}{2}}} \Bigg) \Bigg(    \frac{l}{r} u + \psi \Bigg) 
    +\frac{\alpha B}{A^{\frac{1}{2}}} \Bigg( \frac{2l+2}{r}\psi +\psi' \Bigg)
    +\frac{l(l+1)}{r^2}\Bigg(\frac{B}{A}-1\Bigg)A^{\frac{1}{2}}\alpha u ,
\end{equation}
which actually carries the real dynamics of the quantum scalar field. The last term in Eq.~\eqref{pi mode evolution} is only strictly regular if $A(t,0)=B(t,0)$. Similar regularity conditions were encountered in the case of the classical collapse as well. There, the use of $\lambda$ ensured regularity and hence one can use it here as well. Thus, the final dynamical equation to be used for the scalar mode functions is given by
\begin{equation}
    \dot{\pi}=\partial_r \Bigg( \frac{\alpha B}{A^{\frac{1}{2}}} \Bigg) \Bigg(    \frac{l}{r} u + \psi \Bigg) 
    +\frac{\alpha B}{A^{\frac{1}{2}}} \Bigg( \frac{2l+2}{r}\psi +\psi' \Bigg)
    +\frac{l(l+1)}{r}\frac{\alpha B}{A^{\frac{1}{2}}}\lambda u.
\end{equation}
We emphasize that here we have suppressed the $k,l$ mode indices for $u$, $\psi$ and $\pi$, that is, the above equation represents an infinite set of c-number equations, which however are decoupled as we work with a free scalar field. To get a good approximation for the quantum scalar field, one should evolve as many of these mode functions as possible, which makes the semiclassical simulations much more expensive computationally than the classical ones.

In order to find suitable initial conditions for the quantum mode functions, let us recall that the equation of motion in Minkowski space for $\Phi^{(M)}$ is  $\partial^\mu \partial_\mu \Phi^{(M)} =0$, which is solved by
\begin{equation}
    \Phi^{(M)}=\frac{k}{\sqrt{\pi \omega}}e^{-i\omega t} j_l(kr),
\end{equation}
with $j_l(kr)$'s being the spherical Bessel functions. This means that the solution for $u$ in Minkowski space is given by
\begin{equation}
    u^{(M)}=\frac{k}{\sqrt{\pi \omega}}e^{-i\omega t} \frac{j_l(kr)}{r^l}.
\end{equation}
For simplicity, we shall choose this as our basis of functions at $t=0$. This entails the following initial conditions at $t=0$
\begin{equation} \label{initialcondsquantum}
\begin{split}
    u^0&=\frac{k}{\sqrt{\pi \omega}} \frac{j_l(kr)}{r^l}, \\
    \psi^0&=\frac{k}{\sqrt{\pi \omega}} \Bigg(\frac{j_l'(kr)}{r^l} - \frac{l j_l(kr)}{r^{l+1}} \Bigg), \\
    \pi^0&=-i\omega \frac{k}{\sqrt{\pi \omega}} \frac{j_l(kr)}{r^l}, \\
\end{split}
\end{equation}
since initially we have $A=B=\alpha=1$.

\subsection{Coherent state}
After setting up the evolution of the field operators, one also needs to specify the quantum state of a system.  We choose the quantum state to be a spherically symmetric coherent state
\begin{equation}
\label{coherentsdef}
    \ket{\chi}=\exp{-\frac{1}{2}\frac{1}{\hbar c^2}\int dk \abs{z(k)}^2} \exp{\frac{1}{\hbar c^2}\int dk z(k) \hat{a}^{\dag}_{k,0,0}} \ket{0},
\end{equation}
which is an eigenstate of the lowering operator, $\hat{a}_{k,0,0} \ket{\chi} = z(k) \ket{\chi}$, and $\hat a_{k,l,m}|0\rangle=0$. The first exponential in the $\ket{\chi}$ definition above is fixed by normalisation $\bra{\chi}\ket{\chi}=1$. Coherent states are known to be useful to connect quantum systems to classical ones, and they are ``minimal uncertainty states'' (see {\it e.g.}~\cite{Sanders_2012}), which are in a sense ``the closest to the classical states''. This specific choice of the quantum state will be crucial in this formalism, and it will allow for a direct comparison between the semiclassical collapse and classical collapse.

Let us define the expectation value of the quantum scalar field operator in the chosen coherent state to be
\begin{equation}
    \phi (t,r) \equiv \bra{\chi} \hat{\Phi} (t,r) \ket{\chi}.
\end{equation}
which is often called the ``classical field''. Indeed, we may identify it with the scalar field in the classical gravitational collapse. To see this, note that, making use of the coherent state definition $\hat{a}_{k,l,m}\ket{\chi}=\delta_{l,0}\delta_{m,0} z(k) \ket{\chi}$, we can get
\begin{equation}
    \phi (t,r) = \frac{1}{2 \sqrt{\pi}} \int dk [z(k) \Tilde{u}_{k,0} (t,r) + h.c.].
\end{equation}
where $h.c.$ stands for the Hermitian conjugate of the previous terms and the fraction is due to the fact that $Y^0_0= {1}/{2 \sqrt{\pi}}$. By the evolution equation of the mode function $\Tilde{u}_{k,0} (t,r)$, we see that this expectation value $\phi(t,r)$ satisfies the same equation of motion as the classical scalar field in Section \ref{section2}.
Hence the coherent state choice allows us to separate the classical and quantum contributions in the stress-energy tensor, which is crucial to tap into the established methods to simulate classical gravitational collapse. This is discussed in the next subsection.

\subsection{Quantum stress-energy tensor}
By virtue of the chosen coherent state above, the objects appearing in the stress-energy tensor automatically separate into a sum of contributions from the coherent state expectation value $\phi(t,r)$, which only involves the spherically symmetric mode function $\Tilde{u}_{k,0}(t,r)$, and contributions from the mode functions $\Tilde{u}_{k,l}(t,r)$. Specifically, using Eq.~\eqref{mode_expansion}, the choice of the quantum coherent state Eq.~\eqref{coherentsdef} and some identities for the spherical harmonics, it is straightforward to find that the expectation values for the bilinears appearing in the relevant stress-energy tensor components are
\begin{align} \label{quantum expectations}
    \bra{\chi} \partial_t \hat{\Phi} \partial_t \hat{\Phi} \ket{\chi} &= \partial_t \phi \partial_t \phi + \frac{\hbar c^2}{4 \pi}\int dk \sum_{l=0}^{N_l-1}  (2l+1) \abs{\partial_t \Tilde{u}_{k,l}}^2, \\
    \bra{\chi} \partial_r \hat{\Phi} \partial_r \hat{\Phi} \ket{\chi} &= \partial_r \phi \partial_r \phi + \frac{\hbar c^2}{4 \pi}\int dk \sum_{l=0}^{N_l-1}  (2l+1) \abs{\partial_r \Tilde{u}_{k,l}}^2, \\
 \bra{\chi} \frac{\partial_r \hat{\Phi} \partial_t \hat{\Phi}+\partial_t \hat{\Phi} \partial_r \hat{\Phi}}2 \ket{\chi} &= \partial_t \phi \partial_r \phi 
 \nonumber\\
 &~~~~ + \frac{\hbar c^2}{4 \pi}\int dk \sum_{l=0}^{N_l-1} (2l+1) \frac12(\partial_r \Tilde{u}_{k,l}\partial_t \Tilde{u}^{*}_{k,l} +\partial_t \Tilde{u}_{k,l}\partial_r \Tilde{u}^{*}_{k,l}), \\
    \bra{\chi} \partial_{\theta} \hat{\Phi} \partial_{\theta} \hat{\Phi} \ket{\chi} &= \frac{\hbar c^2}{4 \pi}\int dk \sum_{l=0}^{N_l-1}  \frac{1}{2}(l+1)(2l+1) \abs{\Tilde{u}_{k,l}}^2,
      \label{quantum expectations3}  \\
    \bra{\chi} \partial_{\phi} \hat{\Phi} \partial_{\phi} \hat{\Phi} \ket{\chi} &= \frac{\hbar c^2}{4 \pi}\int dk \sum_{l=0}^{N_l-1}  \frac{1}{2}(l+1)(2l+1) \abs{\Tilde{u}_{k,l}}^2 \sin^2\theta.
    \label{quantum expectationsEnd}
\end{align}
where we have explicitly put back the reduced Planck constant $\hbar$ and the speed of light $c$. Note that here $\phi(t,r)$ is taken to be the same as the scalar field in the classical gravitational collapse. The fact that the quantum expectation values of these operators separate into the classical coherent part and the extra quantum part is merely due to the choice of the coherent state. Note that Eqs. (\ref{quantum expectations3}-\ref{quantum expectationsEnd}) contain only quantum fluctuations, as expected, since the classical pieces of those terms do not depend on $(\theta, \psi)$. In addition, by setting $\hbar$ to zero, one can readily switch off quantum contributions.

Therefore, in our semiclassical simulations, we shall take the decomposed stress-energy tensor quantities $\rho$, $j_A$, $S_A$ and $S_B$ to be
\begin{align} \label{stress1}
    \rho&=n^{\mu}n^{\nu}\langle T_{\mu\nu}\rangle=\frac{1}{2A}\Bigg( \frac{\langle \hat{\Pi}^2 \rangle}{B^2}+\langle \hat{\Psi}^2 \rangle \Bigg) + \frac{1}{Br^2}\langle( \partial_{\theta}\hat{\Phi})^2 \rangle - \frac{1}{2} \mu^2 \langle \hat{\Phi}^2 \rangle, \\
    j_A&=-n^{\mu}\langle T_{\mu r}\rangle=-\frac{\langle \hat{\Pi} \hat{\Psi} \rangle}{A^{\frac{1}{2}}B}, \\
    S_A&=\gamma^{rr}\langle T_{rr}\rangle =\frac{1}{2A}\Bigg( \frac{\langle \hat{\Pi}^2 \rangle}{B^2} +\langle \hat{\Psi}^2 \rangle \Bigg) - \frac{1}{Br^2}(\langle \partial_{\theta}\hat{\Phi})^2 \rangle - \frac{1}{2} \mu^2 \langle \hat{\Phi}^2 \rangle , \\
    S_B&=\gamma^{\theta\theta}\langle T_{\theta\theta} \rangle = \frac{1}{2A}\Bigg( \frac{\langle \hat{\Pi}^2 \rangle }{B^2} - \langle \hat{\Psi}^2 \rangle \Bigg) - \frac{1}{2} \mu^2 \langle \hat{\Phi}^2 \rangle \label{stress4},
\end{align}
where we have defined $\langle \hat{O} \rangle \equiv \bra{\chi}\hat{O}\ket{\chi}$ and used the following relations for the dynamical fields
\begin{equation} \label{stresstensor}
\begin{split}
    \langle \hat{\Phi}^2 \rangle =  \langle  \hat{\Phi} \hat{\Phi} \rangle  ,~~~
    \langle \hat{\Pi}^2 \rangle = \frac{AB^2}{\alpha^2} \langle \partial_t \hat{\Phi} \partial_t \hat{\Phi} \rangle   ,~~~
    \langle \hat{\Psi}^2 \rangle =  \langle \partial_r \hat{\Phi} \partial_r \hat{\Phi} \rangle  ,\\
    \langle \hat{\Pi}\hat{\Psi} \rangle =  \frac{1}{2} \frac{A^{\frac{1}{2}}B}{\alpha} \langle \partial_r \hat{\Phi} \partial_t \hat{\Phi}+\partial_t \hat{\Phi} \partial_r \hat{\Phi} \rangle  ,~~~~
    \langle (\partial_{\theta}\hat{\Phi}^2) \rangle =   \langle \partial_{\theta} \hat{\Phi} \partial_{\theta} \hat{\Phi} \rangle.
\end{split}
\end{equation}

\subsection{Regularisation} \label{regularisation}

The semiclassical Eq.~\eqref{Gqeq} needs regularisation, since the quantum expectation values on the right hand side diverge, which manifests itself as the infinite sums over the mode functions appearing in Eqs.~(\ref{quantum expectations}-\ref{quantum expectationsEnd}). In our lattice simulations, these divergences appear as large, unphysical numbers in the evolution equations that need to be regularised, as is to be expected for a quantum field theory. In a lattice simulation the  popular regularisation scheme used in analytic perturbative computations, dimensional regularisation, can not be used. Instead, we shall introduce Pauli-Villars ghost fields \cite{paulivillars} to cancel the divergences. 

The Pauli-Villars regularisation scheme consists of adding some extra auxiliary (ghost) fields to the physical one, which  by construction cancel the UV divergences. This happens by giving the ghost terms in the Lagrangian the appropriate signs, so that the divergences disappear in the stress-energy tensor. These ghost fields must have larger masses than the physical field, in order to only influence the UV physics. The ghost fields are chosen also to cancel each others' contributions to the various divergences.

Note that even though our quantum field will live on a curved spacetime, it is sufficient to fix the masses of the ghosts by first regularising in Minkowski spacetime. This is because the divergences are UV in nature, hence short distance. In short distances, due to local flatness, the Minkowski approximation still holds. Thus, once the UV divergences are fixed in Minkowski, they are cured on curved spacetimes as well. 

To illustrate the divergences involved in the system, let us take a massless scalar field in Minkowski spacetime. The Lagrangian is just
\begin{equation}
{\cal L} = \frac{M_P^2}2 \int d^4x \Big [ -\frac12 \partial_\mu\Phi\partial^\mu \Phi  \Big].
\end{equation}
This scalar field may be expanded in terms of mode functions similarly as in Eq.~\eqref{mode_expansion}. Then, the energy density and pressure for the field can be found to be \cite{Martin:2012bt}
\begin{equation}
\begin{split}
    \varrho &= \frac{1}{4\pi^2}\int_0^{\mathcal{M}} k^3 dk, \\
    P &= \frac{1}{4\pi^2}\int_0^{\mathcal{M}} \frac{k^3}{3}dk,
\end{split}
\end{equation}
where $k$ is just the wave number and a cut-off $\mathcal{M}$ has been introduced. These integrals are straightforwardly evaluated to be
\begin{align} \label{divergences}
    \varrho &= \frac{1}{16\pi^2} \mathcal{M}^4, \\
    P &= \frac{1}{16\pi^2}\frac{1}{3} \mathcal{M}^4.
     \label{divergences2}
\end{align}
One can see that these terms are divergent, and need to be regularised. The cosmological constant $\Lambda$ can only cancel divergences of the form $\varrho_{div} = - P_{div}$, and so is not able to cure the divergence in Eqs.~(\ref{divergences}-\ref{divergences2}). 

Thus, one needs to introduce Pauli-Villars ghost fields to cancel the divergent density and pressure contributions in Eqs.~(\ref{divergences}-\ref{divergences2}). Let us see how introducing one ghost field $\mathcal{G}_1$ with mass $m_1$ (which is non-zero) changes these expressions. The ghost field is introduced in the Lagrangian with opposite sign kinetic and mass terms, so that 
\begin{equation}
{\cal L} = \frac{M_P^2}2 \int d^4x \Big [ -\frac12 \partial_\mu\Phi\partial^\mu \Phi + \frac12\partial_\mu\mathcal{G}_1\partial^\mu \mathcal{G}_1 +\frac12 m_1^2\mathcal{G}_1^2   \Big].
\end{equation}
Then, the energy density and pressure for the system is
\begin{equation}
\begin{split}
    \varrho_{total} &= \frac{1}{4\pi^2}\int_0^{\mathcal{M}}\Big [ k^3 - k^2 \sqrt{k^2+m_1^2}\Big ]dk, \\
    P_{total} &= \frac{1}{4\pi^2}\int_0^{\mathcal{M}} \Big [\frac{k^3}{3} -\frac{k^4}{3\sqrt{k^2+m_1^2}}\Big ]dk.
\end{split}
\end{equation}
Hence the large $\mathcal{M}$ expansion of these integrals is modified to be
\begin{align} 
    \varrho_{total} = \frac{1}{16\pi^2} \Bigg[&\cancel{\Big(\mathcal{M}^4 -\mathcal{M}^4\Big)}- m_1^2\mathcal{M}^2 \notag \\
    &-\frac{1}{2} m_1^4 \ln\Big(\frac{m_1}{\mathcal{M}}\Big)-\frac{1}{8}(1-4\ln(2))m_1^4 + \mathcal{O} \Big(  \frac{m_1^6}{\mathcal{M}^2} \Big)
    \Bigg], \label{div3}\\
    P_{total} = \frac{1}{16\pi^2} \Bigg[&\cancel{\Big(\frac{1}{3} \mathcal{M}^4-\frac{1}{3} \mathcal{M}^4\Big)} +\frac{1}{3} m_1^2\mathcal{M}^2 \notag \\
    &+ \frac{1}{2} m_1^4 \ln\Big(\frac{m_1}{\mathcal{M}}\Big)+\frac{1}{24}(7-12\ln(2))m_1^4 + \mathcal{O} \Big(  \frac{m_1^6}{\mathcal{M}^2} \Big)
    \Bigg]. \label{div4}
\end{align}
The original divergence coming from the physical field is cancelled, however, due to the nonzero mass of the ghost field, there are new, quadratic, divergences introduced. In addition, now there are finite terms in the expansions from the ghost field, which are unphysical. Therefore, both the new divergences and the finite ghost contributions need to be cancelled.

Notice that the third terms in Eqs.~(\ref{div3}-\ref{div4}) can be cancelled by the cosmological constant $\Lambda$, so that
\begin{align} \label{div5}
    \varrho_{total} = \frac{1}{16\pi^2} \Bigg[&\cancel{\Big(\mathcal{M}^4 -\mathcal{M}^4\Big)}- m_1^2\mathcal{M}^2 \notag\\
    &+\cancel{\Big[\frac{1}{2} m_1^4 \ln\Big(\frac{m_1}{\mathcal{M}}\Big)+\Lambda \Big]}-\frac{1}{8}(1-4\ln(2))m_1^4 + \mathcal{O} \Big(  \frac{m_1^6}{\mathcal{M}^2} \Big)
    \Bigg], \\
    P_{total} = \frac{1}{16\pi^2} \Bigg[&\cancel{\Big(\frac{1}{3} \mathcal{M}^4-\frac{1}{3} \mathcal{M}^4\Big)} +\frac{1}{3} m_1^2\mathcal{M}^2 \notag\\
    &- \cancel{\Big[\frac{1}{2} m_1^4 \ln\Big(\frac{m_1}{\mathcal{M}}\Big)+\Lambda \Big]}+\frac{1}{24}(7-12\ln(2))m_1^4 + \mathcal{O} \Big(  \frac{m_1^6}{\mathcal{M}^2} \Big)
    \Bigg]. \label{div6}
\end{align}
However, we are still left with the quadratic divergences (the second terms) and the unphysical finite ghost contributions (the fourth terms). To cancel these, we need to introduce more Pauli-Villars fields. On the other hand, to make sure that the first terms in Eqs.~(\ref{div5}-\ref{div6}) keeps being cancelled, one needs a total number of fields that is even and therefore two Pauli-Villars fields will not suffice. Note that the second terms will be cancelled by the cosmological constant no matter the number of Pauli-Villars fields.

Hence, one needs to introduce at least three Pauli-Villars fields $\mathcal{G}_1, \mathcal{G}_2$ and $\mathcal{G}_3$, with masses $m_1, m_2$ and $m_3$. To cancel the $\mathcal{M}^4$ terms, one simply uses the opposite sign for the kinetic and mass terms in the Lagrangian for two of them, e.g. $\mathcal{G}_1$ and $\mathcal{G}_3$. To cancel the $m^4$ and $m^2 \mathcal{M}^2$ contributions, the masses must obey the equations
\begin{equation}
\begin{split}
    m_2^2 &= m_1^2 + m_3^2, \\
    m_2^4 &= m_1^4 + m_3^4.
\end{split}
\end{equation}
The solutions to this set of equations always involve $m_1$ or $m_3$ being zero, however, the Pauli-Villars masses must be larger than the physical mass, therefore these are not suitable solutions. In order to obtain a set of masses that are nonzero, two more Pauli-Villars fields have to be introduced, $\mathcal{G}_4$ and $\mathcal{G}_5$, with $\mathcal{G}_5$ having the opposite sign kinetic and mass terms. Thus, the masses must obey
\begin{equation}
\begin{split}
    m_2^2 + m_4^2 &= m_1^2 + m_3^2+m_5^2, \\
    m_2^4 + m_4^4 &= m_1^4 + m_3^4+m_5^4.
\end{split}
\end{equation}
There are many solutions to this set of equations, and we choose the following:
\begin{equation}
    m_2=\sqrt{3}m_1, \quad m_3=m_1, \quad m_4=\sqrt{3}m_1, \quad m_5=\sqrt{4}m_1.
\end{equation}
Therefore, using five Pauli-Villars fields with the above particular masses, the system is regularised. Note that, together with the physical field, this means one has to evolve six dynamical quantum fields now. The Lagrangian for the full regularised semiclassical system is then:
\begin{equation} \label{fulllagrangian}
{\cal L} = \frac{M_P^2}2 \int d^4x \sqrt{-g} \Big [ \frac{R}2 -\frac12 \partial_\mu\Phi\partial^\mu \Phi  - \sum_{i=1}^5 (-1)^{i} \Big( \frac12 \partial_\mu {\cal G}_i\partial^\mu {\cal G}_i +\frac12 m_i^2 {\cal G}^2_i \Big)  \Big],
\end{equation}
where now three ghost fields (with $i=1,3,5$) have opposite sign kinetic and mass terms.
Hence, the system with the Lagrangian in Eq.~\eqref{fulllagrangian} produces regularised stress-energy tensor components and so the expressions in Eqs. (\ref{stress1}-\ref{stress4}) will also be finite. 
\section{Simulation setup}\label{section5}
We shall solve the coupled, highly nonlinear differential equations of the quantum field plus gravity system numerically by discretising the spacetime and putting the equations on a lattice. In this section, the numerical scheme of the spacetime grid is presented along with techniques to solve the discretised system. A subsection is also devoted to the initialisation of the quantum field, which involves choosing a specific number of quantum modes that build up the full quantum operator. It is illustrated how the stress-energy tensor components change for different numbers of modes.

\subsection{Numerical methods}
In the simulation we use a uniform spatial grid consisting of 500 points with $dr=0.025$ and $dt={dr}/{4}$. For spatial derivatives, tenth order finite difference methods are used, and similarly a tenth order implicit Runge-Kutta method \cite{impRK1} is used for time integration. Such high order numerical methods are necessary due to the evolution equations, such as Eq.~\eqref{pi mode evolution}, containing terms involving the $1/r$ and/or $1/r^2$ factor, which are numerically unstable due to the coordinate singularity at $r=0$ if not treated carefully. 

To this end, one also needs to use some artificial dissipation, which acts as damping for the numerical errors. In our code, this is done by adding Kreiss-Oliger terms \cite{dissipation} to the evolution equation of each field. For example, for $\phi(t,r)$, at time step $n$ with the original evolution step denoted schematically by $G(\phi^n)$, $\phi$ at the next time step would be calculated by
\begin{equation}
    \phi^{n+1}=\phi^n + dt \;G(\phi^n).
\end{equation}
Then, adding the extra dissipation term, the evolution equation $G(\phi^n)$ is modified as
\begin{equation}
    G(\phi^n) \longrightarrow G'(\phi^n) = G(\phi^n) - \epsilon (-1)^N dr^{2N-1} \partial_r^{2N} \phi^n,
\end{equation}
where $\epsilon$ is a positive constant which is smaller than one, and $2N$ is the order of dissipation, and is an integer. This added term essentially damps the modes with wavelength close to the grid spacing $dr$. 

In order to preserve the tenth order accuracy of the system, the dissipation term must be at least twelfth order. Even though the accuracy of the system is maintained to be high, in our experience the simulation is more robust when lower order dissipation is used.
Thus, following Alcubierre's notation \cite{alcubierre}, the Kreiss-Oliger term is fourth order with $\epsilon=0.5$ for the quantum mode functions and $\epsilon=0.1$ for the metric fields and $\phi(t,r)$. 
Thus, we use the tenth order methods to stabilise the evolution around $r=0$, but our overall accuracy is fourth order, which is sufficient for our purposes.  

The semiclassical simulation starts with the initial conditions described in subsection \ref{initialcondsclass} with the coherent state expectation value of the quantum field replacing the classical scalar field. In addition, the initial quantum mode functions are described in Eqs.~\eqref{initialcondsquantum}. We also run the classical simulation with the same initial conditions for a comparison. We solve the constraint equations only once at $t=0$ and then use the Hamiltonian constraint to monitor the accuracy of the our simulations.

Note that the accuracy and stability of the system are dependent on the initial coherent state expectation value of the quantum field. This is parameterised by the Gaussian configuration parameters $a, D$. In terms of stability, the most sensitive part of the simulation is the quantum mode functions, specifically the mode functions with high $k$ and $l$ values. These dynamical variables   have the highest tendency to develop instabilities at the origin. Even though both the spatial derivatives and time iteration method are tenth order and dissipation is added artificially to the system, instabilities still arise for high $k$ and $l$ modes, which limits the number of $k$ and $l$ modes that can be added to achieve better accuracy. 
The artificial dissipation, on the other hand, greatly delays the appearance of the instabilities.

One also needs to choose numerical parameters for the mode functions and the integrals, appearing in the stress-energy tensor components, that become discrete sums. Firstly, the mass $m_1$ of the ghost field $\mathcal{G}_1$ must be chosen, which is simply taken to be $m_1=1.0$ in Planck units through our simulations. Crucially, this needs to be larger than the energy of the initial scalar field, which is satisfied by the above mentioned initial conditions. In addition, the continuous integrals of the mode functions over $k$, in e.g. Eqs.~\eqref{quantum expectations}, must be numerically approximated. To this end, a minimum wave number $k_{min}$ is chosen, which will then be the minimal step between wave numbers $dk=k_{min}$, and the integrals over $k$ then become summations multiplied by $dk$. In our simulations we choose $k_{min}=\pi/15$ throughout. Lastly, one needs to choose a finite number of mode functions to involve in the simulation, namely the number of $k$-modes and $l$-modes, which will be denoted by $N_l$ and $N_k$ respectively.  This will be discussed in the next subsection.

\subsection{Vacuum} \label{sectionvacuum}

As mentioned, a crucial choice in the system is the number of quantum modes $\Tilde{u}_{k,l}$ used, which essentially defines the quantum field operator. In practice, we can only use a finite number of quantum modes and generally the more quantum modes one uses, the better simulated the quantum operator is. We find that the maximum number of quantum modes one can use is $N_l=N_k=150$ in our current numerical setup. This limit arises from the amplitude of the modes becoming  too small, saturating the double precision limit of the simulations. 

Due to the finite amount of mode functions, the quantum field is well-defined only in a limited spatial region around $r=0$. In addition, thanks to instabilities from the high $k$ and $l$ valued mode functions, one can simulate the well-defined quantum field only for a finite amount of time also. Hence, there are both spatial and temporal limitations of the simulation. More specifically, for a particular set of initial conditions for the coherent state expectation value, there is a characteristic spatial region where the quantum field is accurately modelled, as well as a characteristic time frame during which all the relevant evolution takes place and through which all evolved mode functions must stay stable. Thus the initial coherent state expectation value must be chosen so that it vanishes outside the region of well-defined quantum field operator, as well as the evolution taking place before the mode functions become unstable. This ensures that everything stays physical in the chosen spatial region and time frame.

Let us discuss the case when the coherent state expectation value is just the vacuum, $i.e.$, $z(k)=0$. Since the coherent state expectation value does not contain divergences, but only the quantum mode contributions do, if the quantum mode contributions are well-defined, then the full quantum operator is as well. The field operator in the vacuum contains exactly these quantum mode contributions only. Since all stress-energy tensor components need to vanish in a well-defined quantum vacuum, this is a good consistency check.  When this is true, we expect that the full quantum field operator is well-defined. 

\begin{figure}[ht]
    \centering
    \includegraphics[width=0.9\textwidth]{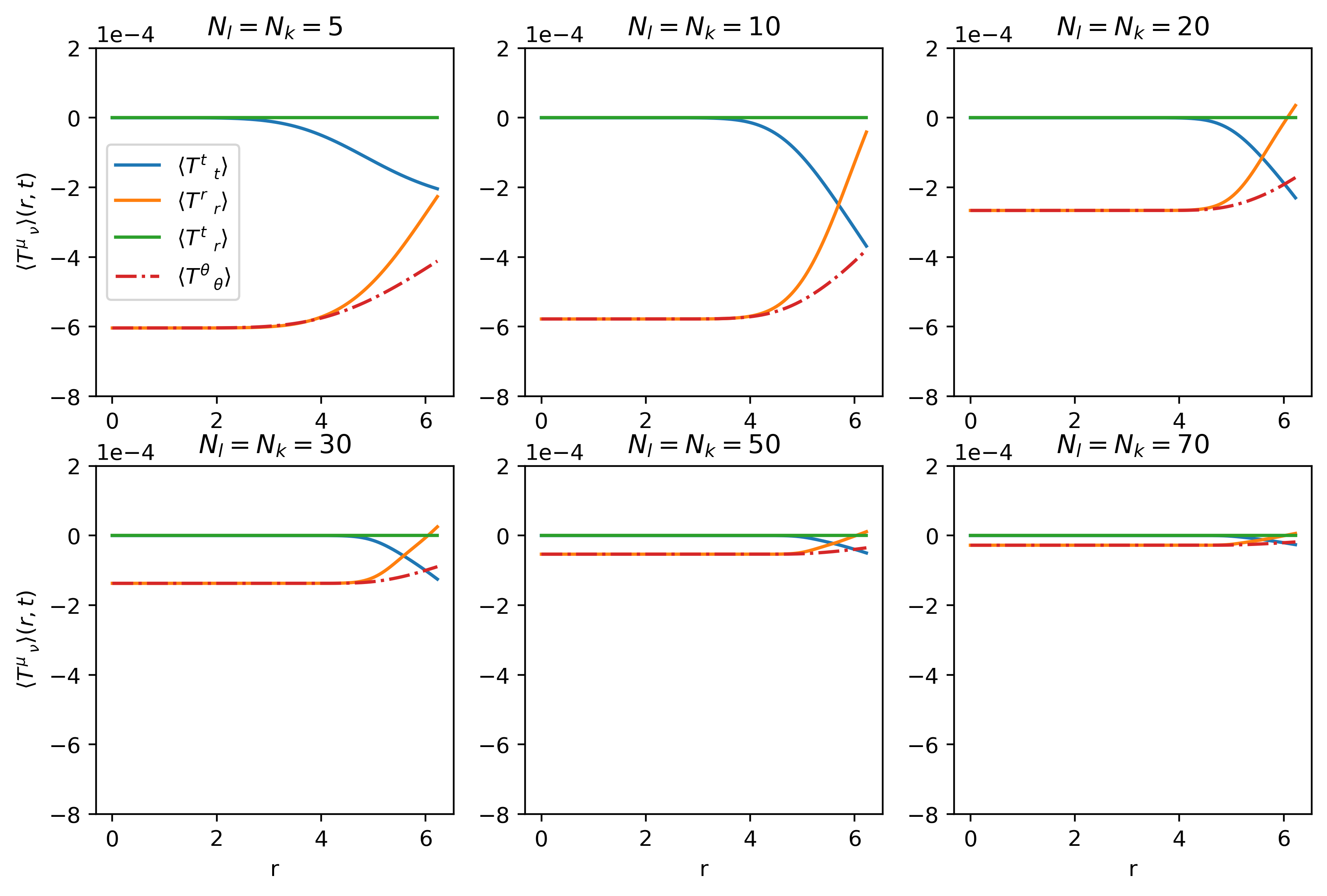}
    \caption{Expectation values of the stress-energy tensor components in the case of zero coherent state expectation value for various numbers of quantum mode functions included, with equal number of $l$-modes and $k$-modes.}
    \label{fig:vacuum}
\end{figure}

As mentioned before, using a larger number of mode functions, one achieves a well-defined quantum vacuum in a larger spatial region. To illustrate this, the stress-energy tensor components, for the case of zero coherent state expectation value, are presented in Figure \ref{fig:vacuum} for various numbers of mode functions with $N_k=N_l$. 
The stress-energy tensor components have been regularised in the previously presented manner. The cosmological constant has been chosen such that the $T^t{}_{t}$ component is zero at $r=0$. This is merely an ad-hoc choice, which then introduces a small deviation of the $T^r{}_{r}$ and $T^{\theta}{}_{\theta}$ components from zero, as seen in Figure \ref{fig:vacuum}. The relative difference between the components is unchanged by the choice of the cosmological constant.

From Figure \ref{fig:vacuum}, it is evident that more mode functions make a better quantum vacuum. The spatial region in which all stress tensor components (approximately) vanish, is more or less constant, namely when $r<5$ for the chosen set of initial conditions. The deviation from zero of the $T^r{}_{r}$ and $T^{\theta}{}_{\theta}$ components, on the other hand, greatly decreases as the number of mode functions is increased. This deviation introduces a small systematic error in our simulation, but generally the errors coming from other sources of the full evolution are actually greater than this error. 

\begin{figure}[ht]
    \centering
    \includegraphics[width=0.9\textwidth]{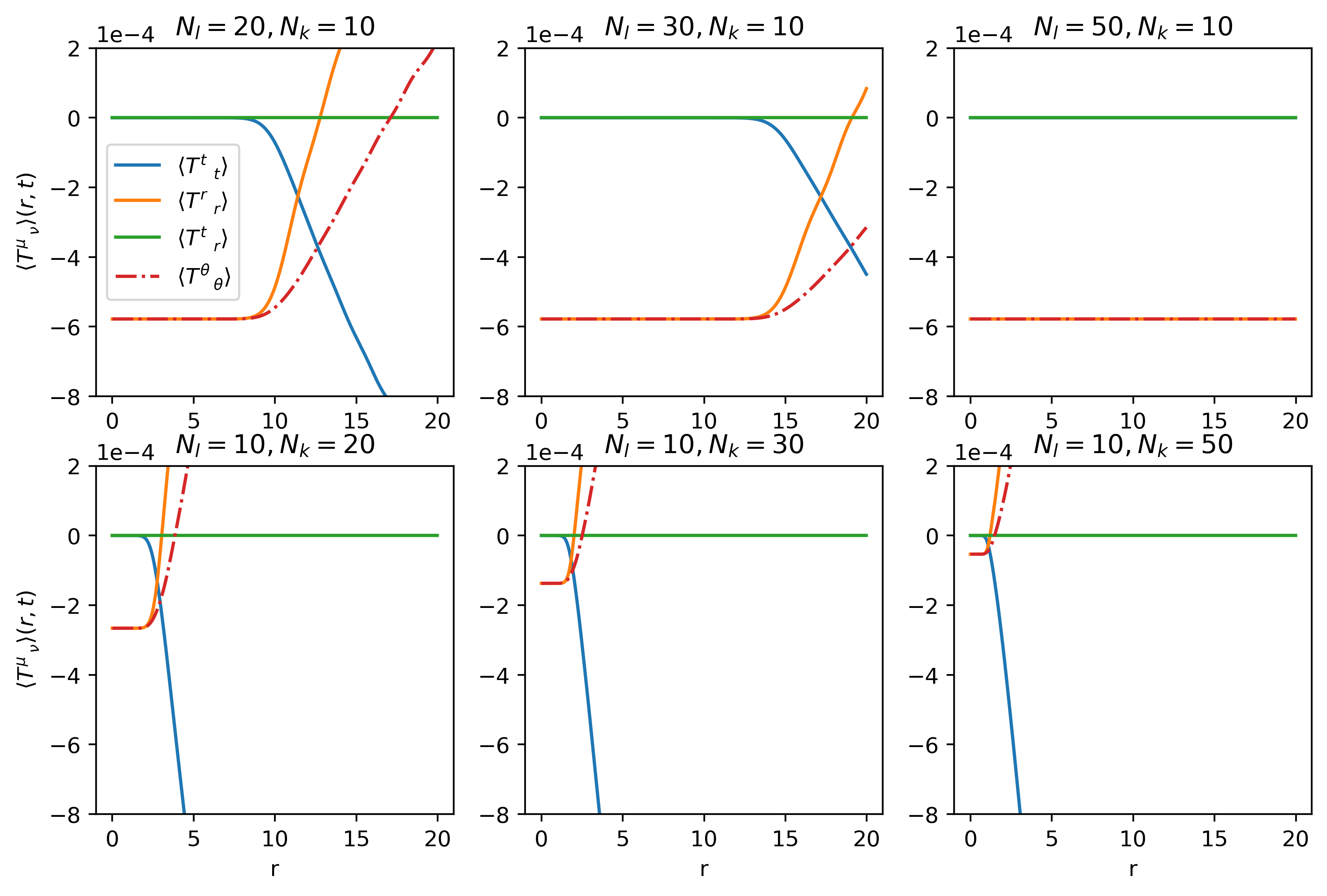}
    \caption{Expectation values of stress-energy tensor components in the case of zero coherent state expectation value for various numbers of quantum mode functions, with different numbers of $l$-modes and $k$-modes.}
    \label{fig:vacuum2}
\end{figure}

Another set of plots are shown in Figure \ref{fig:vacuum2}, where the number of $l$-modes and $k$-modes are not equal ($N_k\neq N_l$), in order to demonstrate how varying one and holding the other number constant influences the vacuum. Note that in these plots, the spatial region is increased in preferable cases, compared to Figure \ref{fig:vacuum}, now $r$ stretching out to $r_{max}=20$. In the first row, $N_k$ is held constant, whilst $N_l$ is increased. The region where the vacuum is well-defined becomes larger as this happens. The relative difference between the components stays constant throughout. In the second row, we hold $N_l$ constant but increase $N_k$. We see now that the relative difference between the components decreases, as one adds more $k$ mode functions. Unfortunately, the flat region where the quantum field is well-defined is actually shrinking as well. This is the reason why, in Figure \ref{fig:vacuum}, even though both the number of $l$-modes and $k$-modes are increased, the flat region stays relatively constant. 

For our purposes, we have managed to simulate a quantum field with $N_l=N_k=50$, {\it i.e.},  2500 modes, for a sufficiently long time in a sufficiently large spatial region to capture the gravitational collapse to a black hole  as well as Choptuik scaling. 

Having established the physical region of the quantum field operator, a ``good'' set of initial conditions for the coherent state expectation value is presented in the next section which vanishes outside the aforementioned spatial region, and its relevant evolution finishes before the mode functions spoil the simulation.

\section{Gravitational collapse and Choptuik scaling}\label{section6}
Once the initial conditions are determined for both coherent state expectation value $\phi(t,r)$ and quantum modes $\Tilde{u}_{k,l}$, we are ready to simulate the gravitational dynamics with a quantum field. In this section, results are shown for initial conditions ending up collapsing into a black hole and also dispersing to infinity. In addition, the system is studied close to the critical amplitude in order to demonstrate Choptuik scaling in critical collapse. We also compare with the classical collapse and identify the quantum deviations in these phenomena.

\subsection{Black hole formation}
After establishing that the quantum field is well-defined in a sufficient region around $r=0$, one can add a nonzero initial coherent state expectation value, and evolve it. The evolution can have two distinct final spacetime structures: Minkowski or Schwarzschild, depending on the initial conditions. We will show the evolution of a chosen initial coherent state expectation value for both cases of black hole formation and no black hole formation. 

\begin{figure}[ht]
\centering
\subfigure[Evolution of $\langle\hat\Phi\rangle$ for no gravitational collapse]{\includegraphics[width=0.49\linewidth]{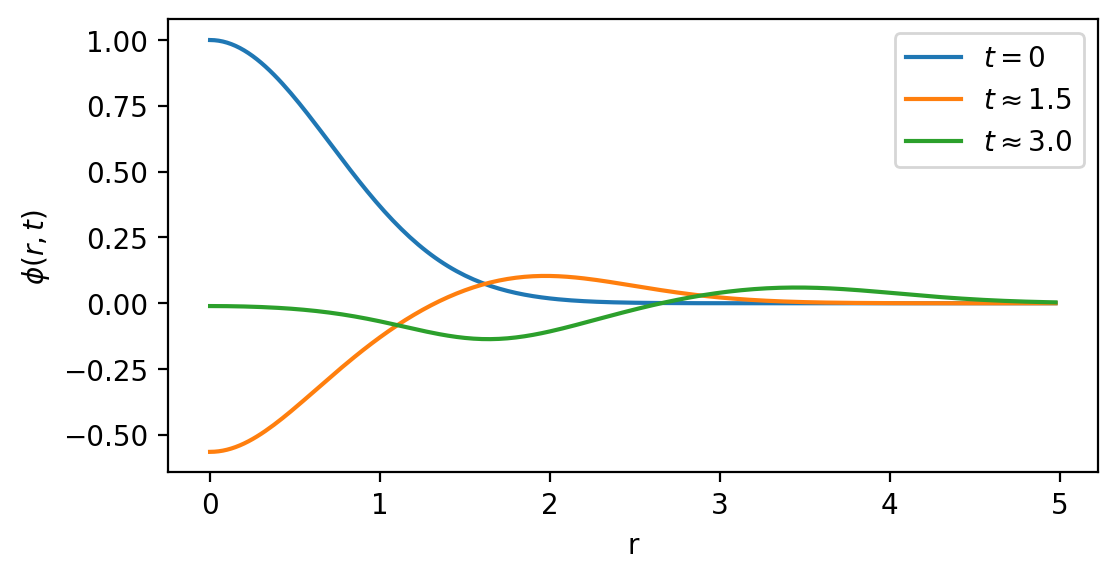} 
} \label{fig:phi1}
\subfigure[Evolution of $\langle\hat\Phi\rangle$ for gravitational collapse]{
  \includegraphics[width=0.44\linewidth]{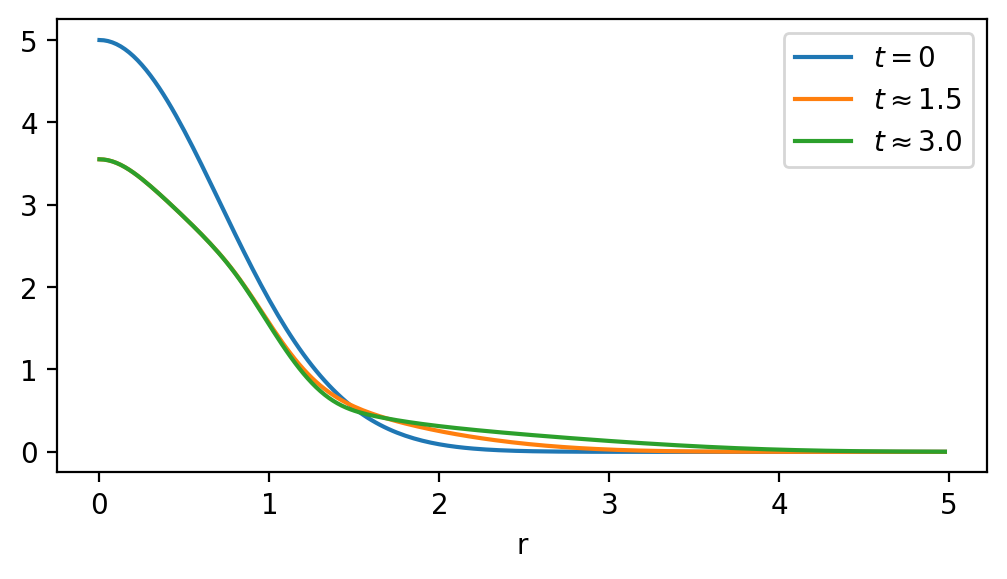}  
} \label{fig:phi2}
\subfigure[Evolution of $\alpha(r=0)$ for no gravitational collapse]{
  \includegraphics[width=0.47\linewidth]{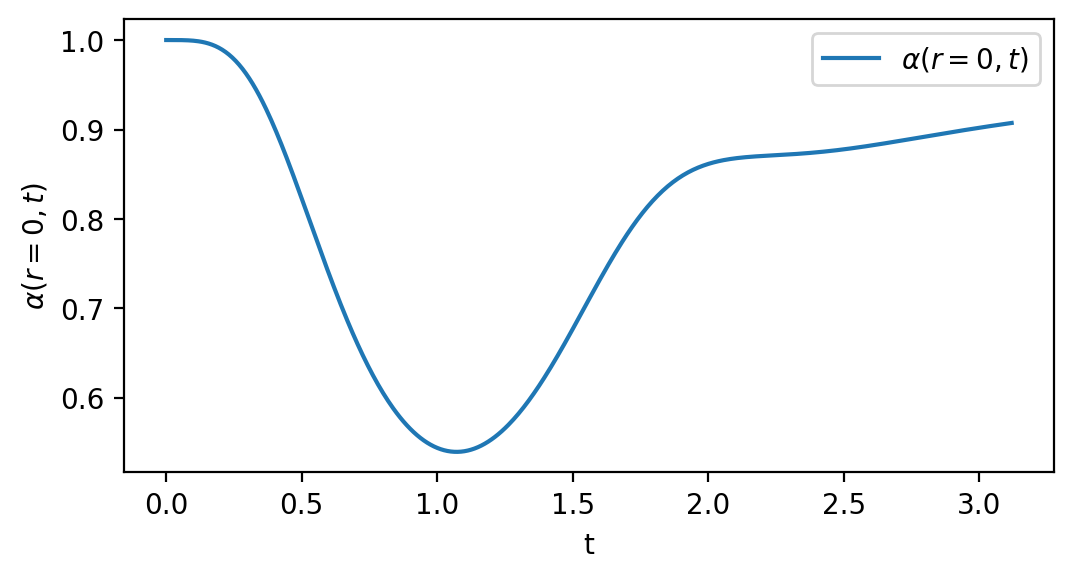}
} \label{fig:alpha1}
\subfigure[Evolution of $\alpha(r=0)$ when a black hole does form, along with the evolution of the apparent horizon area.]{
  \includegraphics[width=0.45\linewidth]{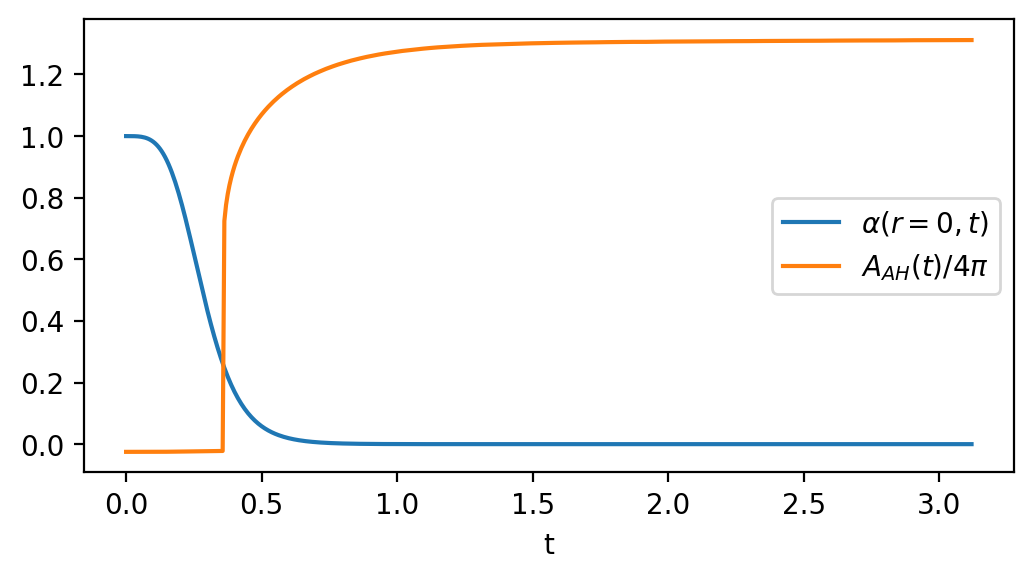}  
}  \label{fig:alpha2}
\subfigure[$\langle \hat{T}^{\mu}_{\: \: \nu}\rangle$ at $t\approx 3.0$ when a black hole does not form]{
  \includegraphics[width=0.48\linewidth]{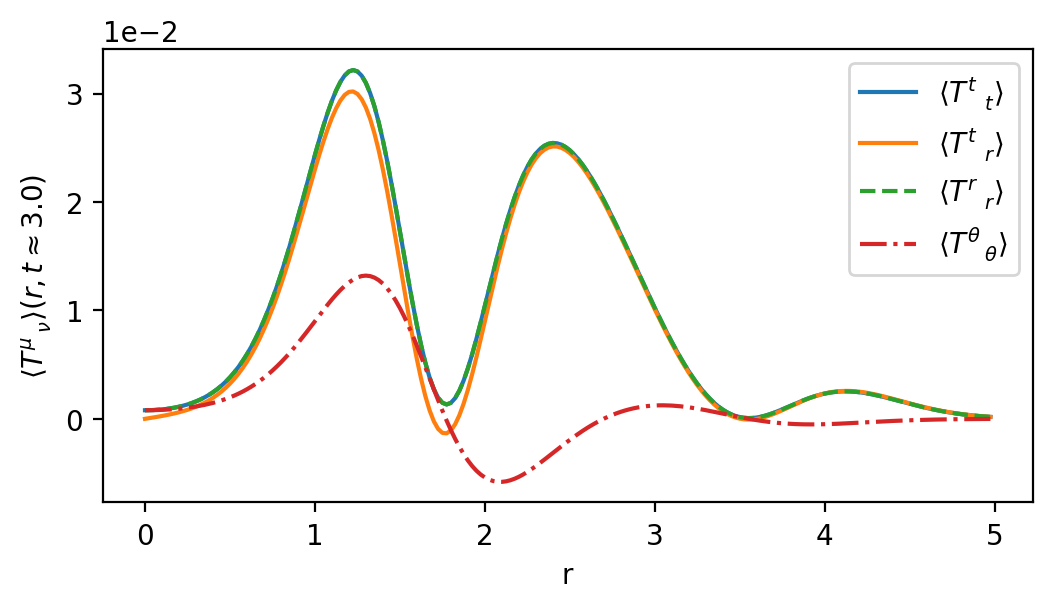}  
}\label{fig:stress1}
\subfigure[$\langle \hat{T}^{\mu}_{\: \: \nu}\rangle$ at $t\approx 3.0$ when a black hole forms]{
  \includegraphics[width=0.46\linewidth]{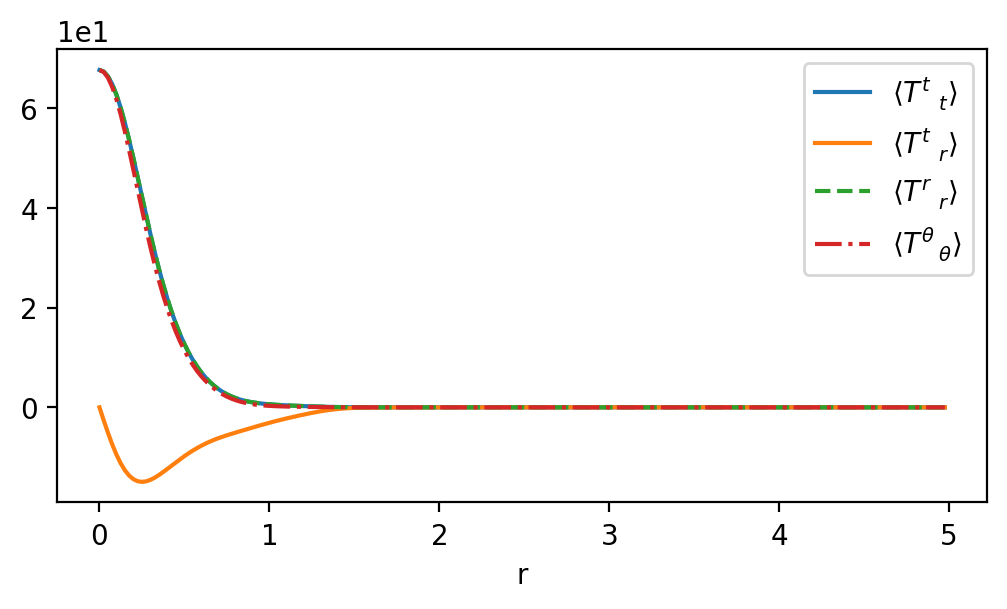}  
}\label{fig:stress2}
\caption{Evolution of wave-packets with initial amplitude $a=1.0$ (left column) and $a=5.0$ (right column). The latter evolution results in gravitational collapse, which can be seen by the collapse of the lapse $\alpha(t,r=0)$ and the appearance of an apparent horizon located at $r_{AH}$.}\label{fig:evolutions}
\end{figure}

As discussed, choosing the coherent state expectation value is equivalent to choosing the initial classical scalar configuration. We focus on an initial classical scalar configuration described by Eq.~\eqref{PhiPsi0} and Eq.~\eqref{initialfield} with $D=1.0$. This initial amplitude can be tuned to achieve an evolution involving a black hole formation or otherwise. For the former case, we can choose $a=5.0$ and for the latter $a=1.0$ is a possible choice. Note that these results were obtained using $N_l=N_k=50$, in other words, 2500 quantum modes. The classical simulations without adding the quantum modes are also performed for a comparison. As expected, for our choices of initial conditions, the classical simulations are a good approximation of the semiclassical ones. More specifically, the plots presented in Figure \ref{fig:evolutions} would be almost indistinguishable from the ones obtained using classical matter. Crucially, though, they are not exactly the same, we will come back to this at the end of this section.

The results for the two separate simulations are presented in Figure \ref{fig:evolutions}. The left column corresponds to the case of no black hole formation (subcritical), and the right column to when the quantum field collapses and a black hole forms (supercritical). (The critical case is presented in the next subsection.) The different rows show various quantities relevant to the evolutions. Let us go through these row by row.

The first row shows plots of the coherent state expectation value of the quantum field, namely $\phi(t,r)=\langle\hat\Phi\rangle$, at three times: at the very beginning, in the middle of the simulation, and at the end of it. In the case of no black hole formation, one can observe that the initial Gaussian wave-packet immediately starts to drop, after which it propagates outward. In the case of the black hole formation (right column), the wave-packet starts a similar evolution, however, it quickly freezes around $r=0$ by virtue of the singularity avoiding gauge choice for the lapse function $\alpha(r,t)$. One can observe a small fraction of the wave-packet still escaping, albeit extremely slowly.

The second row shows the evolution of the lapse function at the very centre, $\alpha(0,t)$. This is significant, since the lapse function vanishes if a black hole forms, due to our gauge choice \cite{alcubierre}. Thus $\alpha(r,t)$ essentially signals the strength of the curvature at any given point. It is plotted at the centre of the spherically symmetric grid, since that is where $\alpha(r,t)$ first reaches zero in case of gravitational collapse. On the left hand side $\alpha(0,t)$ is plotted when no black hole forms. One can see that at its minimal value it reaches about 0.5-0.6, signalling quite a significant curvature. However, it starts increasing and asymptotically approaches one, in other words, Minkowski spacetime is recovering. The plot on the right hand side, corresponding to the supercritical case, shows the evolution of the central value of the lapse function as well as the area of the appearing apparent horizon. The apparent horizon is found using the expansion of the outgoing null geodesics $\Theta$, given by
\begin{equation} \label{nullexpansion}
    \Theta=\frac{1}{\sqrt{A}}\Bigg( \frac{2}{r} + \frac{\partial_rB}{B} \Bigg) - 2K_{B},
\end{equation}
where the physical significance of $K_B$ is that it is the $(\theta, \theta)$ component of the extrinsic curvature. If $\Theta$ vanishes for any given $r$, then there is an apparent horizon located there with radius $r=r_{AH}$ and area $A_{AH}=4\pi r^2_{AH}\left.B\right|_{r_{AH}}$. Thus, in Figure \ref{fig:alpha2}, the central value of $\alpha(r,t)$ is plotted along with $A_{AH}/4\pi$. It can be seen that $\alpha(0,t)$ collapses to zero rather quickly, during which an apparent horizon appears with area $A_{AH}/4\pi \approx 0.8$. The latter then asymptotes to a constant value of $A_{AH}/4\pi \approx 1.25$. The presence of a black hole is unequivocally signalled by these two phenomena; the collapse of the lapse function, and the appearance of an apparent horizon.

The last row in Figure \ref{fig:evolutions} shows the expectation value of the stress-energy tensor components in the coherent state at the end of the simulation, at $t=3.0$. When no black hole forms, as expected, these values are vanishing at $r=0$ and follow the wave-packets propagating outward. On the other hand, in the supercritical case (right figure), all the components of the stress-energy tensor are centered around $r=0$, which is again expected, since most of the energy is located in the black hole around the center. Additionally, note that the stress-energy tensor components are larger by three orders of magnitude in the black hole case compared to the no black hole case.

\begin{figure}[ht]
\centering
        \includegraphics[width=0.99\textwidth]{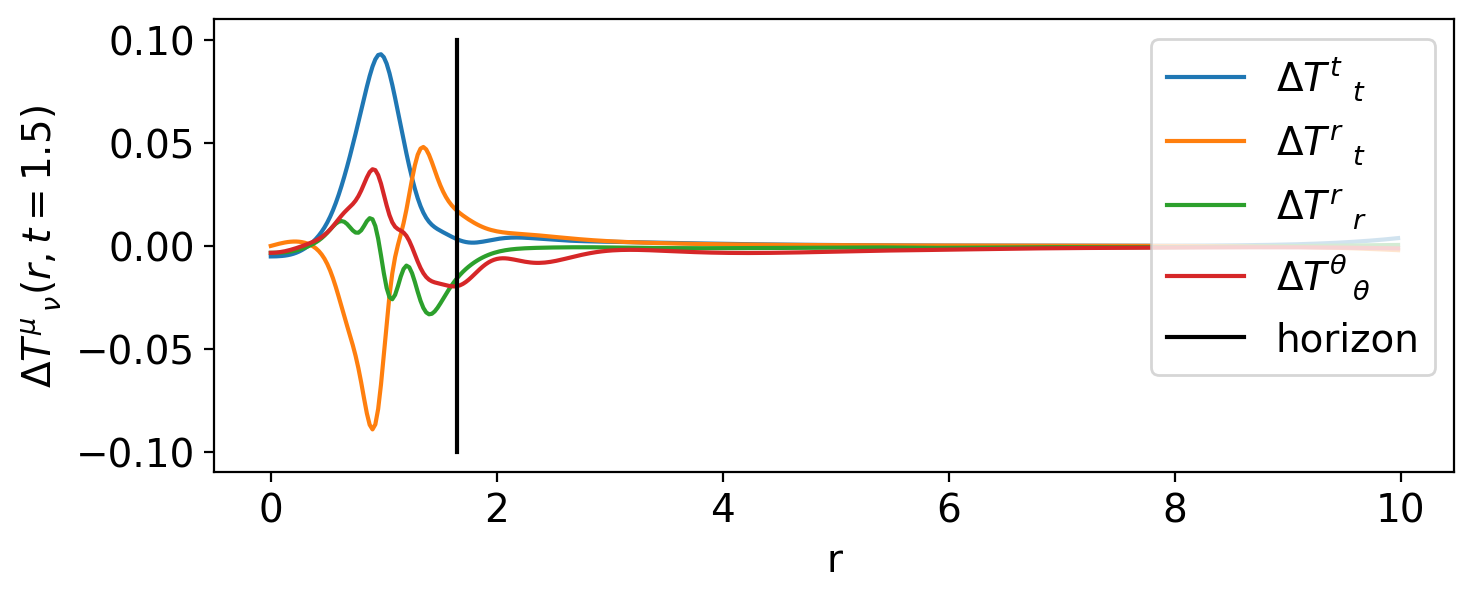}
    \caption{Quantum corrections for the stress-energy tensor components $T^{\mu}{}_{\nu}$, defined by $\Delta T^{\mu}{}_{\nu} = \bra{\chi}\hat{T}^{\mu}{}_{\nu}\ket{\chi} - T^{\mu}_{(c)}{}_{\nu}$ (with $T^{\mu}_{(c)}{}_{\nu}$ from the classical collapse) at $t = 1.5$.}\label{fig:difference}
\end{figure}

Therefore we have seen that the simulation goes as expected for both subcritical and supercritical cases, with clear signs of black hole formation in the supercritical case. These plots, however, are very similar when the matter is just classical as well. The obvious step then is to compare some values in the semiclassical simulation to the classical one. As a proxy to measure the quantum effects, we shall plot the differences between the quantum expectation values of the stress-energy components and the classical stress-energy components:
\begin{equation}
    \Delta T^{\mu}_{\: \: \: \: \nu} = \langle T^{\mu}_{\: \: \: \: \nu} \rangle - T^{\mu}_{(c)\: \: \nu},
\end{equation}
where $T^{\mu}_{(c)\: \: \nu}$ corresponds to the purely classical simulation, without any quantum modes added.
This is plotted in Figure \ref{fig:difference}. Interestingly, as it can be seen in the figure, the largest quantum effects are cloaked by the horizon of the black hole. Nevertheless, there are some quantum deviations spilling out of the horizon. Also note that even though the quantum field is said to be well defined only until $r=5$ (visible in Figure \ref{fig:vacuum}), the quantum mode contributions look flat until around $r=10$. This is just because the quantum mode contributions diverge after the well-defined region ($0<r<5$) rather slowly.

\subsection{Choptuik scaling}

We have explored the subcritical and supercritical evolution of semi-classical gravitational collapse in the previous subsection. The critical behaviour in gravitational collapse has been studied extensively in the classical case, due to the universal scaling phenomena first discovered by Choptuik \cite{choptuik}. A natural question is how the universal scaling behaviour changes in a semiclassical setting. 

\begin{figure}[ht]
\centering
\includegraphics[width=0.48\linewidth]{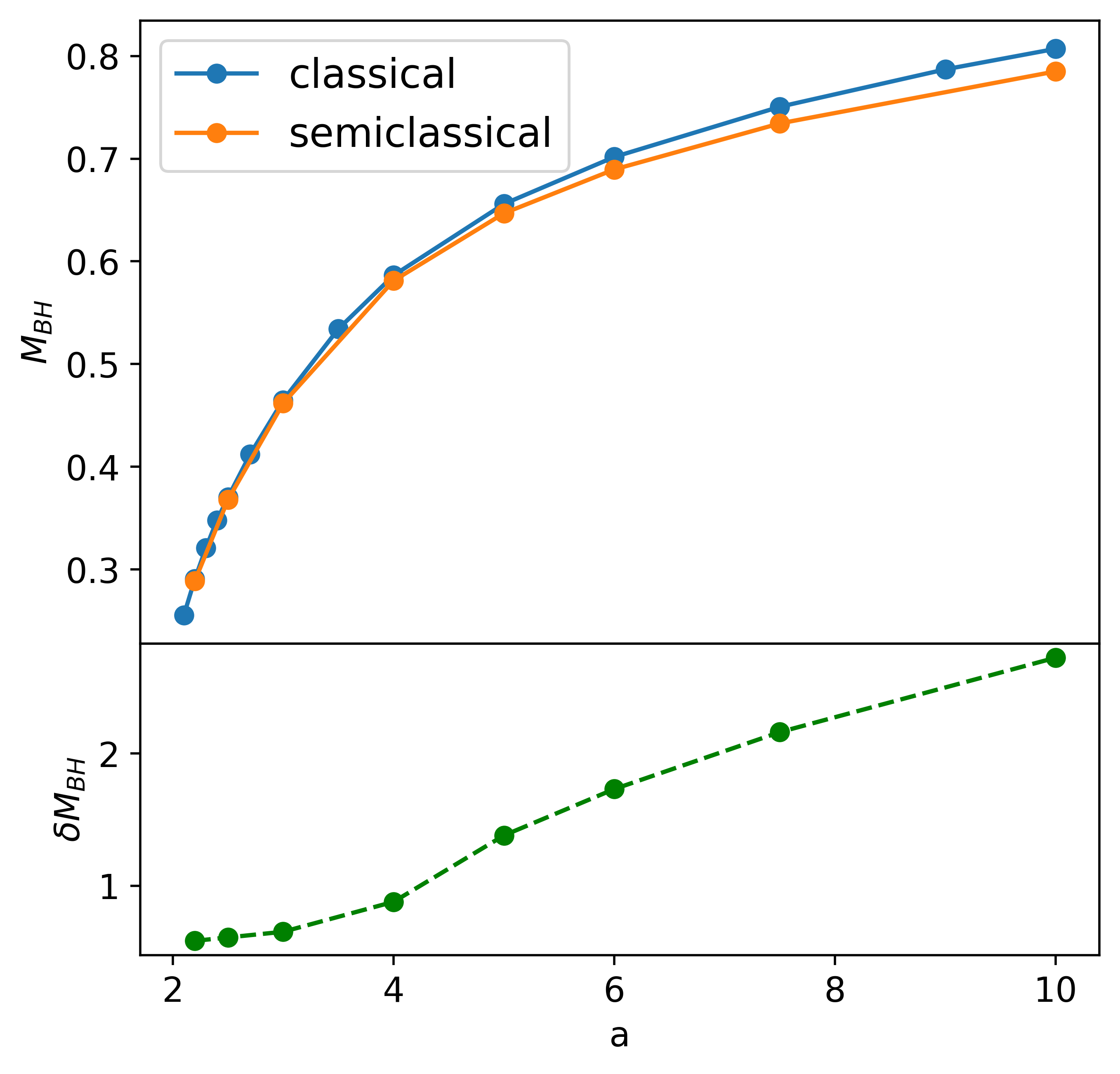} 
\includegraphics[width=0.49\linewidth]{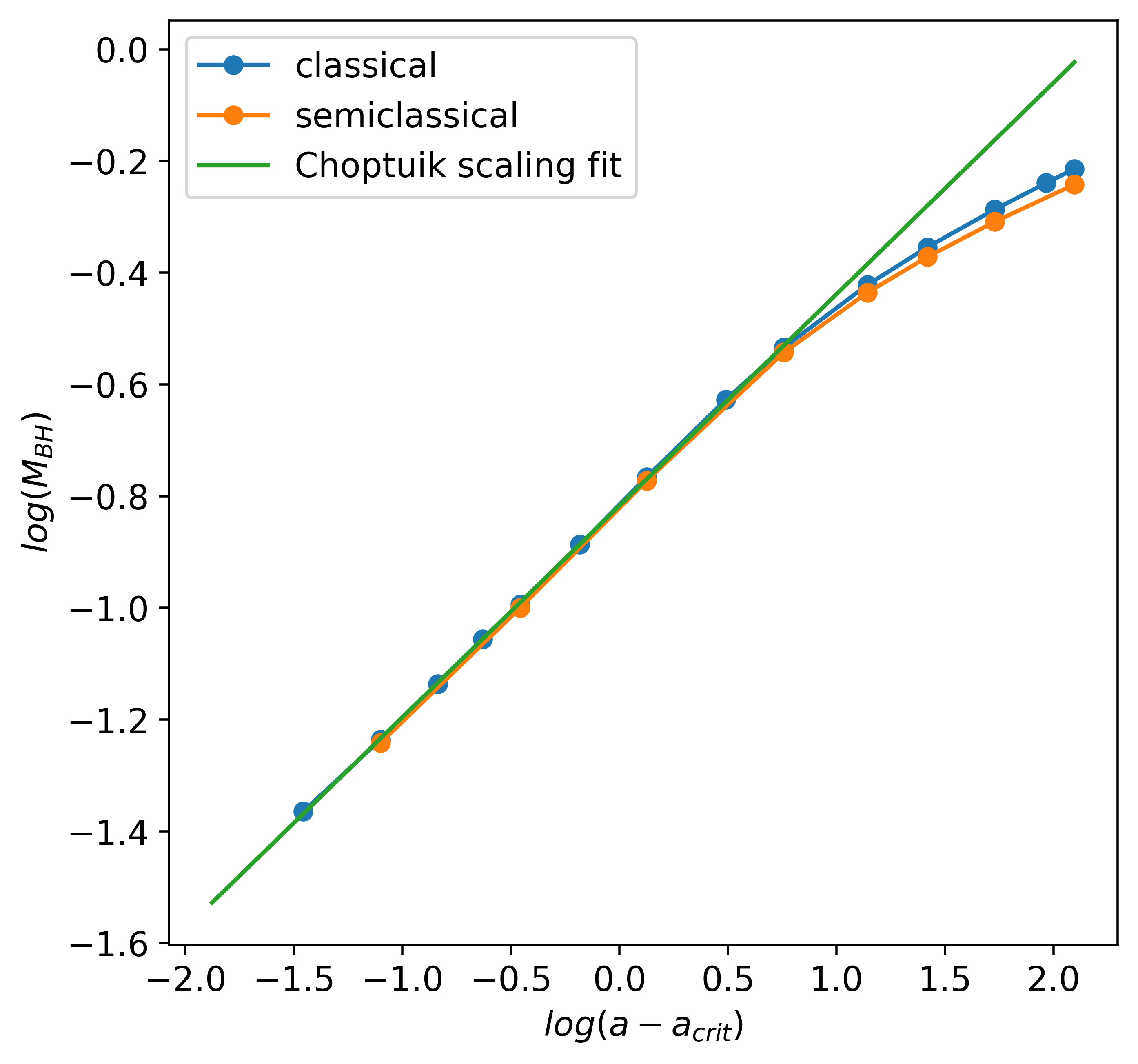} 
\caption{Black hole mass $M_{BH}$ against initial amplitude $a$ for both classical and semiclassical cases. In the left graph, both the $M_{BH}$ and $a$ axis are linear, whereas in the right graph they are plotted (natural) logarithmically to illustrate the proportionality. In addition, the bottom part of the left graph shows the deviation of the semiclassical black hole masses from the classical black hole masses, in percentages. As the system is tuned closer to the critical point, the deviation decreases.}\label{fig:choptuik}
\end{figure}

To answer this question, we run our simulations closer to the critical point, both classically and semiclassically to compare the two. To this end, we fix the parameter $D$ in Eq.~\eqref{initialcondsclass} and vary the initial amplitude $a$. The apparent horizon radius is again found using Eq.~\eqref{nullexpansion}, from which the mass of the black hole can be readily determined by \cite{alcubierre}
\begin{equation}
    M_{BH}=\frac{r_{AH}\sqrt{\left.B\right|_{r_{AH}}}}{2}.
\end{equation}
The results for both the classical and semiclassical simulations using various different initial amplitudes are plotted in Figure \ref{fig:choptuik}.
Note that we have chosen fewer data points for the semiclassical case than the classical case. This is simply due to the computationally expensive nature of the semiclassical simulations. In addition, the difference between the semiclassical and classical black holes can be measured by
\begin{equation}
\label{MBHdef}
    \delta M_{BH}= 100*\frac{M^{(classical)}_{BH}-M^{(semiclassical)}_{BH}}{M^{(classical)}_{BH}},
\end{equation}
which defines the percentages of the deviation in the semiclassical black hole mass from the classical ones. This is also plotted in Figure \ref{fig:choptuik}. One can observe two significant features of the black hole masses in the semiclassical case relative to the classical case. First, the semiclassical black hole mass is always smaller than its classical counter part, which presumably is due to the fact that quantum effects introduce extra dissipation for the matter fields. Secondly, this difference decreases as the black hole masses decrease. However, whether or not this tendency continues as one approaches the critical amplitude even closer is unclear.

Due to the gradients of functions at the centre reaching arbitrarily large values as the critical amplitude is approached, since we are closer to the singularity, the simulation becomes more unstable and the data points are limited in the vicinity of the critical point. 
Nevertheless, the value of the critical amplitude and the critical exponent can be still extracted using curve fitting.
This has been done with the classical points, using a fitting function of the form:
\begin{equation}
    F(a, \gamma, a_{crit}, c)=c (a-a_{crit})^{\gamma},
\end{equation}
where $a$ is the data array of the amplitudes and the other three entries, the critical exponent $\gamma$, the critical amplitude $a_{crit}$ and the constant of proportionality $c$, are parameters of the fit. This function is the expected critical behaviour of the mass of the formed black holes, classically well-known. Using the non-linear least squares fit, the best values of the three parameters are found. Note that in the fitting, not all classical data points in Figure \ref{fig:choptuik} have been used, only the ones with smaller amplitudes, as those are the ones expected to obey the critical behaviour. Indeed, one can see the significant departure from the critical behaviour in case of the last few points (with largest amplitudes). The specific values of the fit plotted in Figure \ref{fig:choptuik} is the following:
\begin{equation}
    \gamma = 0.379 \pm 0.006, ~~~~
    a_{crit} = 1.87 \pm 0.01, ~~~~
    c = 0.442 \pm 0.003.
\end{equation}
The accepted value of the classical scaling exponent in the literature \cite{Gundlach:1996eg, Hod:1996az} is $\gamma=0.374 \pm 0.001$, which is in close agreement with the value found here.      

As mentioned above, the difference between the semiclassical and classical black hole masses decreases for smaller black holes as the system flows to the universal scaling limit.
On the other hand, the curvature at the centre at the critical fixed point is theoretically infinite, so one would naively expect that quantum effects also get larger close to the critical point. Additionally, the strength of Hawking radiation is believed to scale as $M_{BH}^{-2}$, which also suggests this. Whether the semiclassical gravitational collapse approaches the classical one at the critical point is beyond the scope of this paper.

\section{Convergence analysis} \label{sectionconv}
Code validation is important to ensure that the approximate numerical system does indeed converge toward a true exact solution, ruling out the presence of systematic numerical errors that could be falsely identified with physical effects. In this section, the convergence properties of our semiclassical system are analysed in order to validate the simulation numerically. Our simulations have two significant parameters: the number of points used on the physical grid ($N_{grid}$) and the number of mode functions ($N_l,N_k$) used in the semiclassical simulation. These will be analysed separately, since they each individually must be convergent to a true solution for the convergence of the whole semiclassical system.

\begin{figure}[ht]
\centering
   \includegraphics[width=0.49\linewidth]{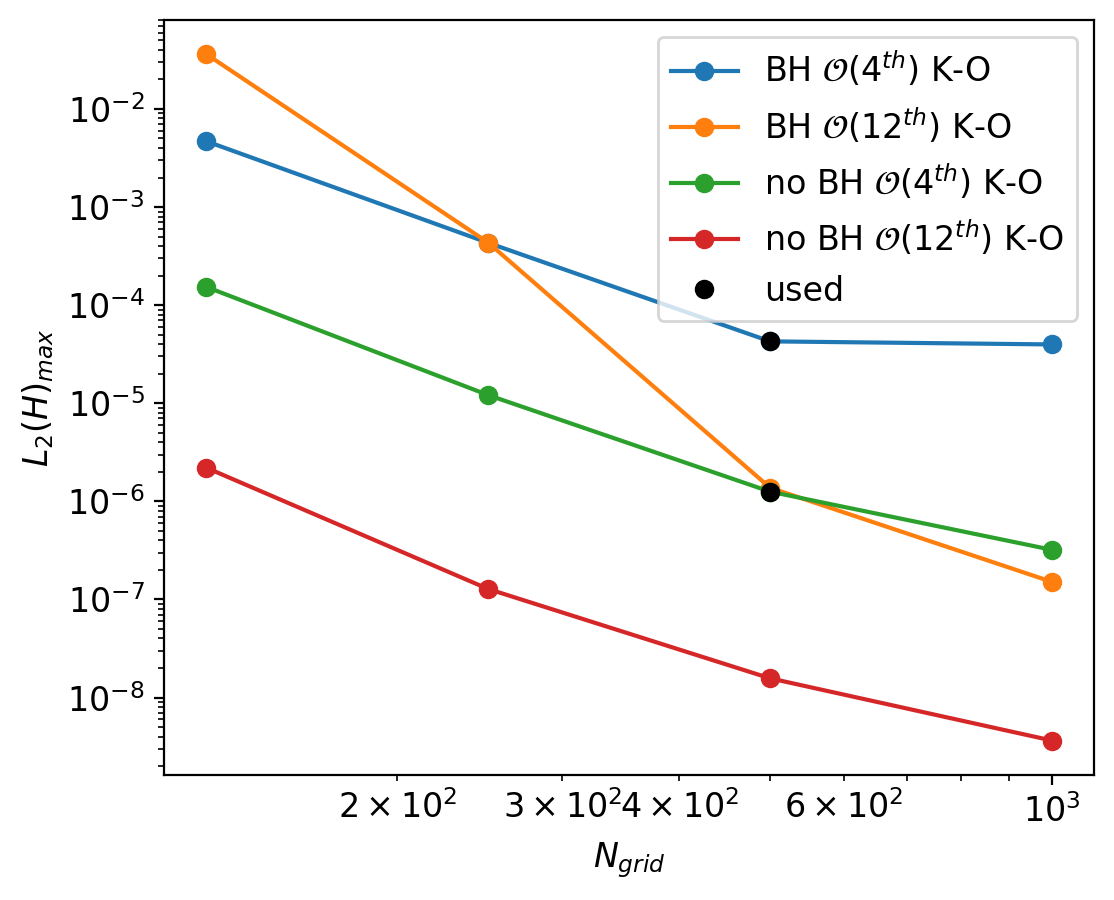}  
   \includegraphics[width=0.49\linewidth]{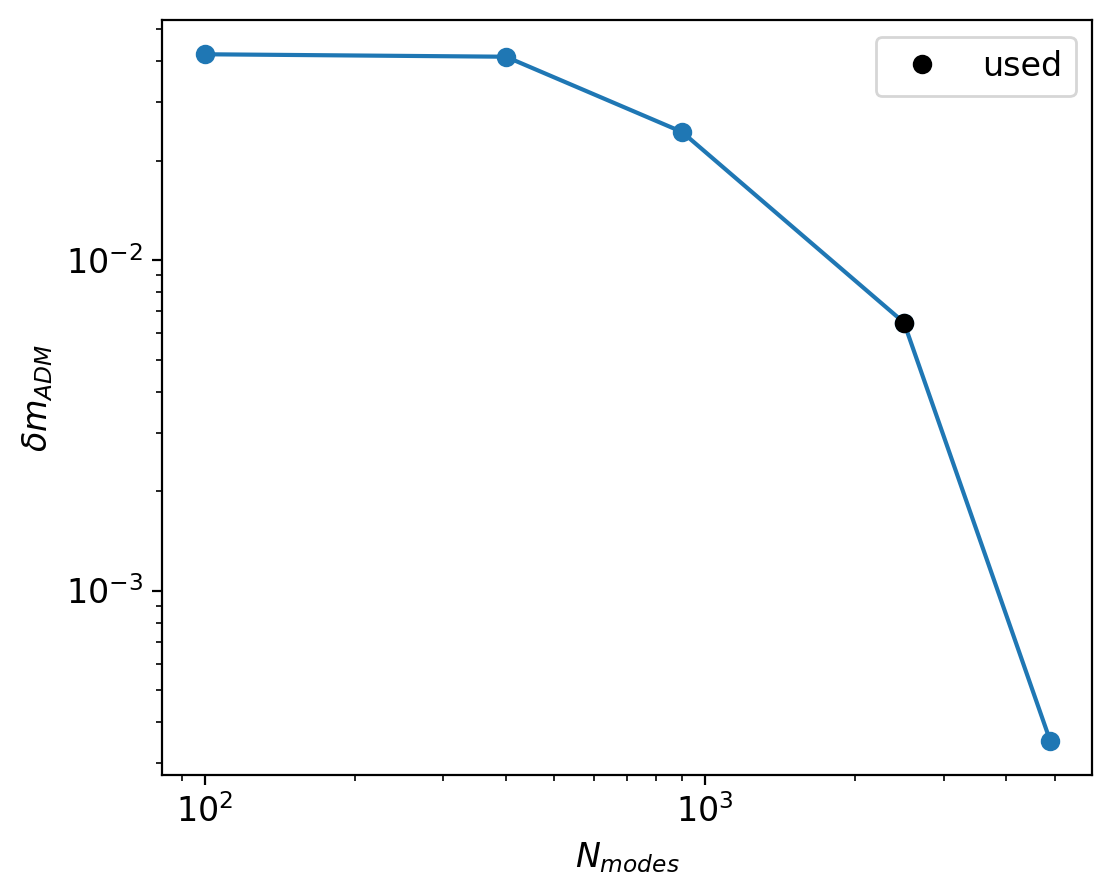}  
\caption{Convergence of the $L_2$ norm of the Hamiltonian constraint with respect to the number of grid points (left) and convergence of the ADM mass with respect to the number of quantum mode functions used (right). The dots labeled with ``used'' are the ones that we use in Fig.~\ref{fig:evolutions} and Fig.~\ref{fig:difference}.}\label{fig:convergence}
\end{figure}

Let us first discuss the number of grid points, $N_{grid}$. Specifically, by this, we mean the number of discrete spatial points and temporal steps that our 2D computational grid is divided into. Thus $N_{grid}$ is inversely proportional with the separation between grid points spatially $dr$ and also temporally (through the relation $dt=dr/4$). Hence, as one uses more grid points for a given physical region, the separation between the points decreases with the grid approaching the smooth continuous spacetime. To check convergence for this parameter, the $L_2$-norm of the Hamiltonian constraint is analysed, which is a staple object in numerical relativity used for such purposes. This is plotted on the left hand side of Figure \ref{fig:convergence}, for four different cases: supercritical evolution with low and high order artificial damping and subcritical evolution with low and high order damping. As the purpose of this convergence study is to calibrate the numerical accuracy of our finite difference method for the grid and the damping term, these simulations are done using classical matter, in other words, without any quantum mode functions added. From the figure, convergence for all cases is clear. It is also indicated on the graph which specific configurations are used to plot Figures \ref{fig:evolutions} and \ref{fig:difference}.

In order to check convergence with respect to the number of quantum modes added, another staple numerical relativity object is chosen; the ADM mass. This is calculated using the so-called ``Schwarzschild-like'' mass \cite{alcubierre}, which converges to the correct ADM mass very rapidly, sufficiently far away from the center of the spacetime where all the mass is contained. The ``Schwarzschild-like'' mass is defined by
\begin{equation}
    m_{Schw}=\sqrt{\frac{A_{sphere}}{16\pi}}\Big( 1-\frac{(dA_{sphere}/dr)^2}{16\pi A(r,t) A_{sphere}}  \Big),
\end{equation}
where $r$ is an arbitrary radial coordinate, $A_{sphere}=4\pi B(r,t) r^2$ is the area of spheres at radial coordinate $r$ and $A(r,t)$ is the $rr$ component in  metric, Eq.~(\ref{metric0}). Then the ADM mass is just
\begin{equation}
    m_{ADM}= \lim_{r\to R} m_{Schw},
\end{equation}
where $R$ is some sufficiently large value of radial distance. Then, to measure the convergence of the system with respect to the number of added quantum modes, we define the deviation from the initially calculated ADM mass as time evolves
\begin{equation}
    \delta m_{ADM} = \frac{m_{ADM}(t)-m_{ADM}(t=0)}{m_{ADM}(t=0)}.
\end{equation}
This is plotted on the right hand side of Figure \ref{fig:convergence} with various different numbers of quantum mode functions added, for the black hole formation case at $t=1.5$ and $r \approx 8.0$. The convergence is clear as the number of modes, $N_{mode}$, is increased. Again, the specific amount of quantum modes used to obtain the results presented in Figure \ref{fig:evolutions} and \ref{fig:difference} is indicated as well.

\begin{figure}[ht]
\centering
        \includegraphics[width=0.85\textwidth]{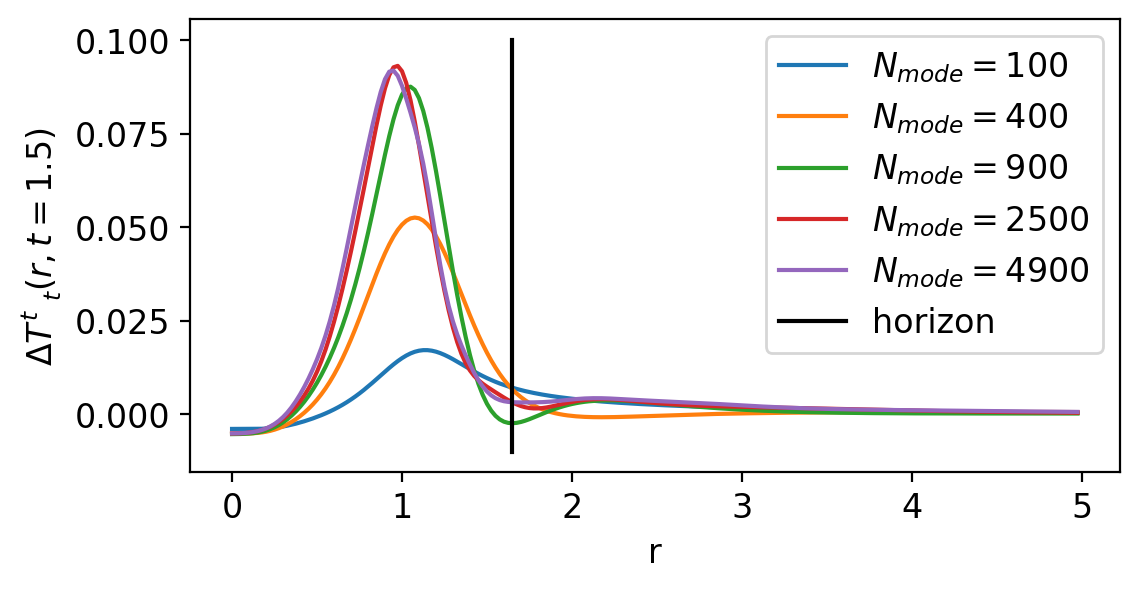}
    \caption{Convergence study of quantum corrections for the stress-energy tensor component $\Delta T^{t}{}_{t} = \bra{\chi}\hat{T}^{t}{}_{t}\ket{\chi} - T^{t}_c{}_{t}$ (with $T^{t}_{c}{}_{t}$ from the classical collapse) at $t = 1.5$.}\label{fig:difference2}
\end{figure}

While monitoring the convergence of the Hamiltonian constraints and the black hole mass provides a useful sanity check for the whole numerical setup, to ensure convergence of Figure \ref{fig:difference} specifically, we plot $\Delta T^t{}_{t}$ for various different numbers of quantum modes in Figure \ref{fig:difference2}. From this, we see that adding even more quantum modes would not qualitatively change the observed quantum deviations for the stress-energy tensor.

\begin{figure}[ht]
\centering
        \includegraphics[width=0.85\textwidth]{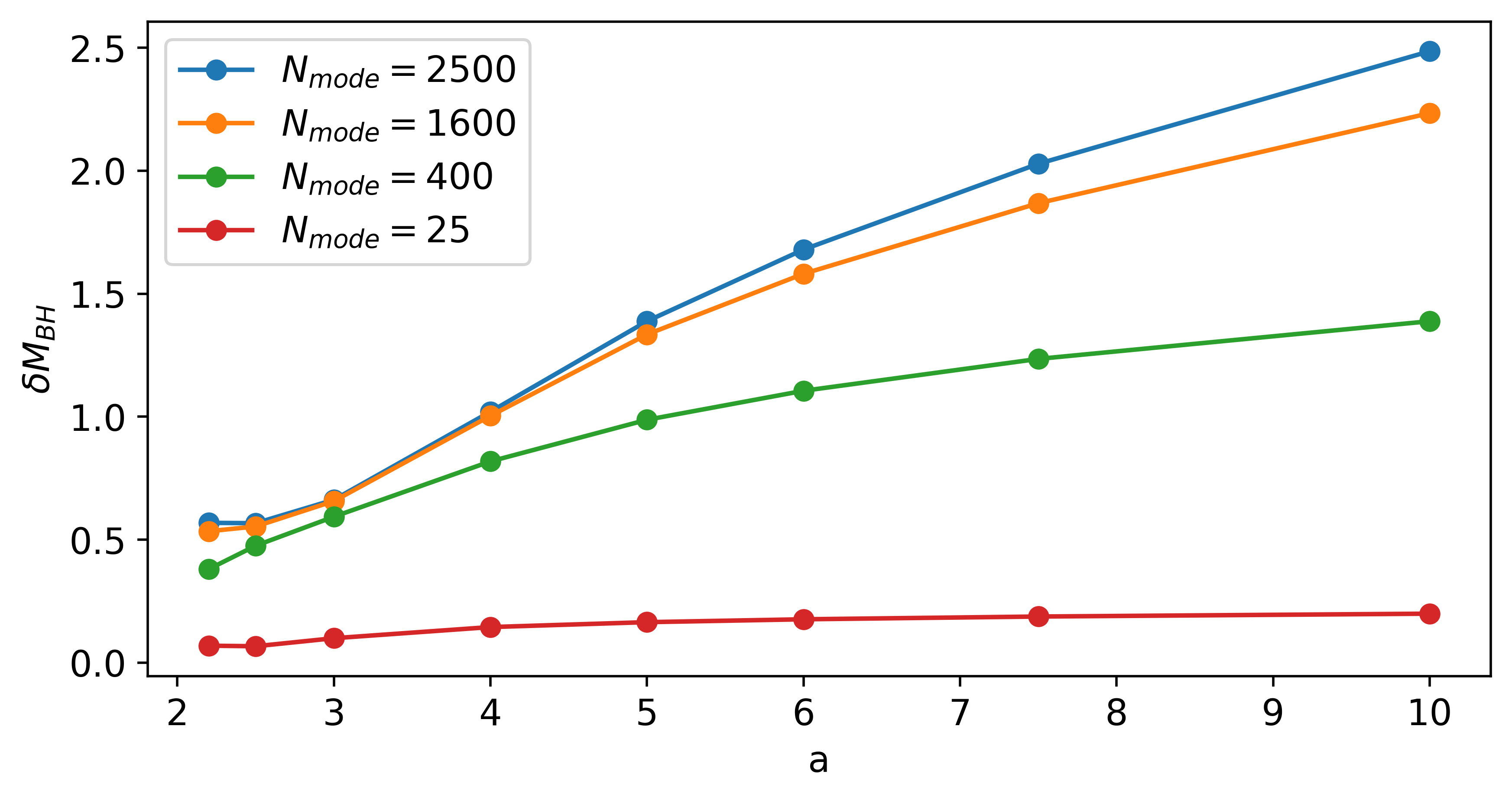}
    \caption{Convergence study of quantum corrections to the black hole mass $M_{BH}$, defined by Eq.~\eqref{MBHdef}. The vertical axis labels the percentage of the quantum corrections.}\label{fig:massdifference}
\end{figure}

We also want to check the convergence of the Choptuik scaling. For this, we shall check the convergence of the black hole mass differences defined in Eq.~\eqref{MBHdef} for Figure \ref{fig:choptuik}. This can be seen in Fig \ref{fig:massdifference}, where we plot the percentages of the black hole mass differences for various numbers of quantum modes. We see that increasing the number of quantum modes makes the black hole mass corrections converge rather quickly. There is little difference between the curves with $N_{mode}=1600$ and $N_{mode}=2500$ ($N_{mode}=2500$ being used in Figure \ref{fig:choptuik}) for small initial amplitude $a$, and good convergence has also been achieved for the larger $a$ in Figure \ref{fig:choptuik}.

\section{Conclusions} \label{sectionsum}
In this paper a new formalism has been introduced to collapse quantum fields into black holes. Coherent states have been utilised in order to relate the semiclassical system to the purely classical one. This allows one to replicate the already well-studied classical simulations in a semiclassical setting, hence giving rise to a convenient framework to study quantum effects in gravitational collapse. 
The simulations have been validated by convergence analysis for the $L_2$-norm of the Hamiltonian constraint and the deviation in the ADM mass during the evolution. In addition, the convergence of the quantum effects themselves in the results has been illustrated.

The formalism has been implemented in numerical simulations to solve the coupled differential equations of the Einstein field equations along with the Klein-Gordon (operator) equation of a fully quantum mechanical massless scalar field in a spherically symmetric spacetime. It has been shown that using a finite number of quantum modes the quantum field operator is well-defined in a finite but sufficiently large spatial grid where interesting dynamics can take place. Results have been presented for initial conditions of both the subcritical and supercritical gravitational collapse, which are in agreement with the expected classical evolution. The presence of quantum effects have been explicitly demonstrated in the case of supercritical collapse by comparing the stress-energy tensor components in the semiclassical simulation to the ones in the corresponding classical simulation. The emerging quantum effects are found to be located around the apparent horizon.

The black hole masses in the semiclassical and classical systems have also been compared. The semiclassical black hole mass seems to be always smaller than its classical counterpart, and also the difference between them grows for increasing initial energy input in the simulation. The scaling behaviour of the black hole mass with the initial amplitude of the input wave-packet was investigated for both classical and semiclassical systems. It is found that they both obey the Choptuik scaling near the critical amplitude. The scaling exponent and critical amplitude have been computed using a least squares fitting, with the results in agreement with the accepted classical values.

There are various avenues of possible future work. Among others, one option is to use this formalism to find numerical evidence for the appearance of Hawking radiation in the supercritical case. 
In terms of the numerical sector, it has been mentioned that the higher order quantum modes introduce some instabilities to the simulations, which means that the semiclassical evolution can only be maintained for a certain time period before the accuracy is spoiled.  While this has not posed existential difficulties for the current work, this might be crucial, for instance, to study the Hawking radiation. Various improvement can potentially be made for the numerical schemes, for example by employing the spectral methods or adaptive mesh refinement. These techniques might sustain stability for the semiclassical simulation for a longer period of time, which would enable a more extensive study of the emerging quantum effects in gravitational collapse.

\acknowledgments

BB is supported by an STFC studentship, PMS acknowledges support from STFC grant ST/P000703/1 and SYZ acknowledges support from the starting grants from University of Science and Technology of China under grant No.~KY2030000089 and GG2030040375, and is also supported by National Natural Science Foundation of China under grant No.~11947301, 12075233 and 12047502, and supported by the Fundamental Research Funds for the Central Universities under grant No.~WK2030000036.

\addcontentsline{toc}{section}{Bibliography}
\bibliographystyle{unsrt}
\bibliography{main2}

\end{document}